# Identifying patterns and recommendations of and for sustainable open data initiatives: a benchmarking-driven analysis of open government data initiatives among European countries

Martin Lnenicka, Anastasija Nikiforova, Mariusz Luterek, Petar Milic, Daniel Rudmark, Sebastian Neumaier, Caterina Santoro, Cesar Casiano Flores, Marijn Janssen, and Manuel Pedro Rodríguez Bolívar

**Abstract**: *Open government and open (government) data are seen as tools to create new opportunities, eliminate or at least reduce information inequalities and improve public services. More than a decade of these efforts has provided much experience, practices, and perspectives to learn how to better deal with them. This paper focuses on benchmarking of open data initiatives over the years and attempts to identify patterns observed among European countries that could lead to disparities in the development, growth, and sustainability of open data ecosystems. To do this, we studied benchmarks and indices published over the last years (57 editions of 8 artifacts) and conducted a comparative case study of eight European countries, identifying patterns among them considering different potentially relevant contexts such as e-government, open government data, open data indices and rankings, and others relevant for the country under consideration. Using a Delphi method, we reached a consensus within a panel of experts and validated a final list of 94 patterns, including their frequency of occurrence among studied countries and their effects on the respective countries. Finally, we took a closer look at the developments in identified contexts over the years and defined 21 recommendations for more resilient and sustainable open government data initiatives and ecosystems and future steps in this area.*

## 1 Introduction

Open data - especially Open Government Data (OGD) - is a critical aspect of open government. They have grown in popularity and there is a trend to increase the understanding of how to go beyond building an Open Data Ecosystem (ODE). The underlying assumption is that a mature and sustainable ecosystem will improve openness, transparency, accountability, public participation, a collaboration between different actors (including G2C, G2G, G2B), along with citizens' trust in government and quality of life moreover the reuse of OGD is expected to drive innovation, (co-)create public/social value, promote economic growth, etc. (Kawashita et al., 2020; Susha et al., 2015; Veljkovic et al. 2014; Zuiderwijk et al., 2021). In addition, the OGD concept contributes to Sustainable Smart Cities, data-driven economy, open innovation, digital twins, or Society 5.0. However, this requires understanding the current state, weaknesses, and strengths of OGD initiatives (or its parts, such as the OGD portal and other platforms providing access to open data) and position, among others.

The position of a country in relation to other countries provokes debates about "*where are we?*", "*where should we be?*" and "*what needs to be done to achieve it?*". Answering these questions has triggered the development of many benchmarks, indices, ranks, and other types of comparisons over the past decade, e.g., see Lnenicka et al. (2022), Meuleman et al. (2022), Susha et al. (2015), Zheng et al. (2020), Zuiderwijk et al. (2021). Also, many study-specific benchmarks proposed a combination of the above or a completely new benchmarking framework or index, e.g., de Juana-Espinosa and Luján-Mora (2019), Kawashita et al. (2020), Kubler et al. (2018), Machova and Lnenicka (2017), Neumaier et al. (2016), Quarati et al., (2021), Vancauwenberghe et al. (2018), Zheng et al. (2020). For more examples, see Lnenicka et al. (2022) and Zheng et al. (2020).

Benchmarking itself is a management method used to "*calibrat[e] [your] efficiency against other organizations, getting the inspiration and building on other peoples experiences*" by "*mak[ing] a comparison between parts of or the entire operation*" (Karlöf, 2003, p. 65). Benchmarking tracks and monitors the progress of the OGD initiative (or a specific part of it), comparing it to competitors to drive improvements and the continuous development of the public data ecosystem at different levels of such as OGD portal and its specific feature, data release and maintenance, created impact, etc. (Dawes et al., 2016; Lnenicka et al., 2022; Meuleman et al., 2022; Welle Donker and van Loenen, 2017; Zuiderwijk et al. 2021). Measuring open government (data) progress provides insight into the strengths and weaknesses of a particular aspect, serving as a powerful incentive/stimulus for scoping improvements, or to accelerate its development.

These benchmarks are intended for practitioners, and academics. They provide a qualitative assessment of the OGD results, the entire e-government and whether the goals are being achieved. In this context, OGD are often viewed as a policy instrument and an e-government tool (Meuleman et al., 2022). Some studies have found a correlation between open government benchmarks and the evolutionary stages of e-government (Veljkovic et al., 2014). At the country-level, benchmarks help to assess whether corrective actions can be initiated to improve effectiveness. The lessons and actions can be learned from others and be used to adopt best practices or to validate actions.

While benchmarks can provide insights, help comparisons with competitors, or serve as a source of inspiration, choosing from the wide variety which ones to consider when setting an agenda for changes or improvements can be difficult. This problem increases due to the rankings and their individual results across different categories of OGD initiatives or artifacts, if they overlap across different indices, may vary significantly (Kawashita et al., 2020; Lnenicka et al., 2022; Meuleman et al., 2022; Susha et al., 2015; Zuiderwijk et al., 2021). Users of the OGD, policymakers, civil servants, and government officials may struggle to decide which benchmark model to apply. At the same time, rarely are existing indices' underlying methodologies fully disclosed in detail, including how data for the index were collected and the specific calculations as reports (Lnenicka et al., 2022). Moreover, in many cases, the methodology used by these benchmarks changes over time to be more aligned with the current state-of-the-art. Still, politicians, researchers, and enthusiasts remain unaware, focusing on the result. This makes it difficult to interpret the results for decision-making or determine future actions and can lead to intentional misinterpretation or manipulation in favor of open data owners, politicians, or policymakers (Bannister, 2007; Lnenicka et al., 2022; Nikiforova and McBride, 2021; Susha et al., 2015; Zuiderwijk et al., 2021).

According to the current research (Lnenicka et al., 2022; Nikiforova and McBride, 2021; Susha et al., 2015; Zuiderwijk et al., 2021), when countries rank high according to a certain benchmark, responsible actors can use this to pursue open data benchmarks rather than considering the actual demands for advancement with open data (Susha et al., 2015). In other words, these results are used as an argument to decrease their efforts to develop further initiatives and ignore other benchmarks where their country ranks lower, neglecting the opportunity to identify measures to improve their progress in the OGD initiative. Meanwhile, Nikiforova and McBride (2021) emphasize that many indices and benchmarks are based on assessments made by experts who evaluate OGD initiatives or examine governments' self-reports. It leads to a situation where governments focus their OGD initiatives and particular artifacts to maximize the scores in these international rankings, which does not necessarily lead to actual improvement. This aligns with Zuiderwijk et al. (2021) critique, according to which existing benchmarks and indices fail to provide insight into weaknesses, why they are considered weaknesses, and what corrective actions can be taken.

Many benchmarking models' issues are caused by the complexity of the open government (data) initiative and the OGD ecosystem, as well as the lack of a standardized evaluation and measure of the progress and evolution of the open government (data) initiative. This results in differences in scope, purpose, underlying assumptions, and definitions of each benchmark model, that also evolve over time even within the same benchmarking model, which makes it necessary to study changes in these indicators/metrics over time (Lnenicka et al., 2022; Meuleman et al., 2022; Zuiderwijk at al. 2021). Such analysis is needed to ensure that changes in the results are due to the success or failure of efforts rather than changes in a particular evaluation model. At the same time, considering the open government (data) phenomenon complexity and the goals expected to be achieved through them, Sandoval-Almazan and Gil-Garcia (2016) postulate the need for an integrated framework to comprehend and evaluate open government and its primary elements. Similarly, Susha et al. (2015) called for investigating the extent to which benchmarks can be integrated to produce a more complete and inclusive open data benchmark. In both cases, however, it is important to remember that open data benchmarks are only approximations of reality (Susha et al., 2015).

This paper aims to identify patterns observed in open data initiatives over the years and evaluate their effects that could lead to disparities and divides in ODEs development and benchmarking of ODEs. To achieve this, we have defined the following Research Questions (RQs):

> RQ1: What are the patterns observed in open (government) data initiatives over the years?
>
> RQ2: What are the effects of the identified patterns that could lead to disparities and divides in the development and benchmarking of ODEs?

To answer the RQs, we need to first examine existing benchmarks, indices, and rankings of open (government) data initiatives, as well as to find the contexts by which these initiatives are shaped, both of which then outline a protocol to determine the patterns. The composite benchmarks-driven analytical protocol is used as an instrument to examine the understanding, effects, and expert opinions concerning the development patterns and current state of ODEs implemented in the countries under study. For this purpose, we applied a 3-round Delphi method to identify, reach a consensus, and validate the observed development patterns and their effects that could lead to disparities and divides. Specifically, this study conducts a comparative analysis of different patterns of open (government) data initiatives and their effects in the eight selected countries using six open data benchmarks, two e-government reports, and other relevant resources, covering the period of 2013–2022.

Based on the results, we provide recommendations regarding the patterns observed, their similarities, and their effect on the development and benchmarking of ODEs to improve the understanding of what needs to be adjusted in open data initiatives. These adjustments can lead to improved performance in applied indices and rankings and, more importantly, will facilitate the achievement of the benefits with which OGD are associated. While this is expected to be important in instructing ODEs' stakeholders (mainly policymakers), the findings will help to identify research gaps to be further explored by researchers, including answering questions set by Susha et al. (2015).

We found that approaches to benchmarking of open data initiatives have been affected by the development of e-government over the years. We can argue that the development of e-government, especially after 2000, influenced how services were implemented within the public sector's internal processes, towards citizens and businesses, and created the basis for open data initiatives and OGD disclosure and reuse. This mainly concerns data infrastructure and related processes, i.e., how public sector agencies and institutions can identify, pre-process, and publish their data on data portals.

The rest of the paper is structured as follows: Section 2 provides a research background, Section 3 presents the methodology of the research, Section 4 deals with data collection and contexts, Section 5 presents the case study, i.e., its summary and findings, including recommendations and future steps, while Section 6 establishes the discussion and elaborates on limitations. The final section concludes the paper.

## 2 Research background

### 2.1 Benchmarking, indices, and rankings – an overview

Index, rank, and benchmarking are terms that are used when referring to approaches to measuring and benchmarking efforts at the level of a unit of analysis, such as a country. According to Lnenicka et al. (2022), index and rank are two related terms, which are often used interchangeably. The term "index" is most often found in the name/title of a particular index. Its main purpose is usually to rank entities such as countries in the case of OGD indices. However, since most indices are explained and detailed in a written report, the terms "benchmark" and "benchmarking" more accurately reflect the purpose of the index.

According to Sammut-Bonnici (2015), benchmarking as a method was first developed in the management sciences. It can be done as an internal, external, or international process, where the latter becomes more popular as digital technologies provide more efficient and effective ways to collect and process data. This is especially valid in open data systems, as it provides opportunities to identify national OGD systems that can become reference points for excellence. As Sammut-Bonnici (2015) argues, during international benchmarking, products, and processes are compared in a global context and at different stages of the life cycle.

Benchmarking can use both simple and composite indicators, where the choice of indicators largely depends on the complexity of the system being evaluated. The more variables needed to describe the system fully, the more effective the use of composite indicators is. For open data systems, simple indicators such as the number of published datasets or the number of visits may be sufficient to parameterize an open data portal. However, for a holistic analysis, it is necessary to go beyond the data on the portal itself and consider contexts such as the use of open data or the impact of their release on the economy or society. Such an analysis will naturally necessitate the use of composite indicators.

A step forward in the direction of shaping the benchmarking instruments for open government initiatives was made by Kawashita et al. (2020), who proposed new dimensions for analysis. Their research resulted in building the Measurement Guide, which utilizes metadata, meta-method, and meta-theory to explain how benchmark models measure various aspects of the OGD. Meanwhile

Nardo et al. (2008) focused on composite indicators which compared and ranked country performance and aimed to provide an improvement in the techniques currently used to build them to improve the quality of their outputs. In their approach, composite indicators should be viewed as a way to encourage debate and stimulate public attention.

Michener (2015) insists that composite indices can potentially captivate public interest because they represent a monolithic strategy for measurement. They are easy to comprehend as they provide a simplified answer to the question: "*how good are we compared to others.*" However, this simplification is often criticized, especially regarding the structure and components of the benchmarking frameworks and how the outputs are validated from a statistical point of view. Bannister (2007) states that as a ranking system needs a final single scale, a method of arriving at such a score must be decided with no fixed or commonly agreed rules. Consequently, if two rankings use the same set of simple indices, their final scores may vary if they assign different weights to those indices.

Lnenicka et al. (2022) suggest that these indices and rankings must be standardized to reflect globalization and the need for transnational cooperation in the open government movement. Even if they are constantly updated in methodologies to follow current trends, their application over time results in incomparable releases of the same index. According to Bannister (2007), it can result from time-sensitive metrics. It is especially valid in the case of OGD systems, as both understandings of the openness and technologies used to achieve it change over time.

To summarize, to better understand open data benchmarking approaches, it is imperative to validate the construct of a composite indicator by verifying whether different dimensions of OGD, measured with the same assessment/score, correlate with each other (González et al., 2017). This approach can assist in determining if various governance dimensions correspond to fundamentally distinct phenomena or aspects of the same thing. The convergent validity of an indicator is tested by comparing the indicator vis-à-vis another indicator that aims to capture a related underlying phenomenon. One way to test the reliability of a construct is to check whether the construct produces consistent results over different periods. OGD efforts rapidly evolve, and the underlying data and methodology have changed. There should be a certain level of consistency in the results.

## 2.2 Benchmarking of open (government) data initiatives

There are several studies conducted in recent years exploring existing open data indices and rankings, benchmarks of the OGD initiatives, and respective reports. Kawashita et al. (2020) explored how the Open Data Charter principles are measured in OGD assessment, coming up with a list of six international OGD assessments, namely the Open Data Inventory (ODIN), Global Open Data Index (GODI), the European Open Data Maturity Assessment also known as Open Data Maturity Report (ODMR), Open Data Barometer (ODB), the Open, Useful and Re-usable data (OURdata) Index, and the Open Data Monitor.

Zuiderwijk et al. (2021) compared methodologies used to measure, benchmark, and rank governments' progress in OGD initiatives. Using a critical meta-analysis approach, the authors compared nine benchmarks - Open Data Readiness Assessment, ODB, GODI, Open Data Economy, ODMR, Open Government Index (OGI), OGD Report (also Organization for Economic Cooperation and Development (OECD) report or OURdata Index), ODIN, and OGD by The Economist Intelligence Unit (EIU) further limiting their study to six OGD benchmarks, namely ODB, ODIN,

ODMR, OGI, OURdata Index, and ODIN. Although the impact of open data is typically not quantified, the study indicates that both the academic open data progress models and the current OGD benchmarks employ quite different measurements and approaches. They grouped the indices into three groups: (1) benchmarks that consider the publication of government as one of the most important characteristics of open data progress looking exclusively at open data publication (GODI, Open Data Economy, OURdata Index, ODIN); (2) benchmarks that exclusively focus on the use or potential use of OGD (WJP Index and EIU); (3) benchmarks that look into both aspects (ODB, ODMR, Open Data Readiness Assessment).

Lnenicka et al. (2022) identified six popular and widely rankings (independently or forming an input to other OGD systems) used – GODI, ODB, OURdata Index, ODIN, ODMR, and the Open Government Development Index (OGDI), which were rigorously inspected by analyzing their underlying methodologies and indicators, and how they have changed over time, and, more importantly, whether the results of different editions of the same index can be comparable and used as the basis for decision-making on the development of specific aspects and input data to determine further actions for the OGD initiative. They grouped the indices into three groups depending on their focus, i.e., what aspect(s) of the ODE they measure: (1) openness of selected data categories (GODI, ODIN); (2) various aspects of the ODE through a (large) number of variables (OURdata Index, ODMR); (3) those that try to combine both of the above approaches (ODB, OGDI).

All of those benchmarking initiatives were introduced during the last decade (see Table 1). The first editions of two open data indices were released in 2013. The first is the Open Knowledge Foundation's (OKF) GODI, which tracks the state of open datasets from the government and how well they adhere to standards that define the openness of data and content. The Second is the World Wide Web Foundation (W3F)'s ODB, which aims to give an overview of best practices for open data. The ODIN by Open Data Watch (ODW), the ODMR by the European Union (EU), and the OURdata Index by the OECD all were released in 2015. The ODIN evaluates key data categories' conformance to open data standards and their coverage and availability. It only considers the information on the National Statistics Offices' official website (NSOs). The ODMR aids European nations in enhancing their open data initiatives. Finally, the availability, accessibility, and reuse of public data serve as the foundation for the OURdata Index, which evaluates government efforts to follow the G8 Open Data Charter. The most recent attempt to benchmark OGD is the OGDI index by the United Nations Department of Economic and Social Affairs (UN-DESA), introduced in 2020. Since all these indices aim to evaluate progress over time, their methodology is constantly revised to reflect modern trends and demands.

*Table 1.* Overview of open data indices and rankings published by international organizations, extended from Lnenicka et al. (2022)

| Title | Publisher | First report | Last report | No. of reports | No. of countries covered by each report |
|---|---|---|---|---|---|
| GODI | OKF | 2013 | 2016 | 4 | 60; 97; 122; 94 |
| ODB | W3F | 2013 | 2017 | 5 | 77; 86; 92; 115; 30 |
| OURdata Index | OECD | 2015 | 2019 | 3 | 30; 34; 33 |

| Title | Publisher | First report | Last report | No. of reports | No. of countries covered by each report |
|-------|-----------|--------------|-------------|----------------|------------------------------------------|
| ODIN  | ODW       | 2015         | 2022        | 6              | 125; 173; 180; 178; 187; 192             |
| ODMR  | EU        | 2015         | 2022        | 8              | 31; 31; 32; 31; 32; 35; 34; 35           |
| OGDI  | UN        | 2020         | 2022        | 2              | 191;193                                  |

GODI - Global Open Data Index, ODB - Open Data Barometer, OURdata Index - Open, Useful and Re-usable data Index, and the Open Data Monitor, ODIN - Open Data Inventory, ODMR - Open Data Maturity Report (also European Open Data Maturity Assessment), OGDI - Open Government Development Index
No. of reports - number of reports / editions published between the first and the most recent report

All of those indices are constructed as composite measures to capture multiple elements and their relationships in the ODE. The highest level of the index is usually represented by a *score*, i.e., a final score obtained by combining several *sub-indices*, *dimensions*, *pillars*, *indicators*, etc., where: (a) *dimension*, also called *sub-index* or *pillar*, represents the various levels and aspects on which the score is built; (b) *indicator*, also called category or component, represents various types of variables, and may have a form of a composite or a simple indicator; (c) *metric*, the lowest level of description, is a simple measure that is represented by a value for each entity (Lnenicka et al., 2022).

Based on the analysis of these indices and ranking, we can see that the ODMR can be considered as the most detailed. It assesses the EU Member States and the candidate or potential candidate countries. It has the most cohesive sample, as all of those countries are modeled towards or aspire to the same system of values on which openness is based. ODMR is also the most continuous benchmark, published annually, launching new editions continuously since 2015. Finally, its methodology is reviewed annually considering the ongoing developments in the field. Although this may negatively affect the ability to analyze the progress over the years, it provides insights into the compliance of the OGD initiative with current trends, including its resilience and sustainability. As it turned out at the later stages of our analysis, it is also most often used when reflecting on the current state of development of open data initiatives and when planning other activities (GODI and ODB were also used in some countries, while they were actively maintained). More precisely, we find that:

    1) GODI, ODB, and OURdata Index are no longer active. One of their goals was to raise awareness of open data benchmarking in its early days, tracking the availability and accessibility of datasets and their degree of openness;

    2) OGDI is a supplementary index derived from one of the three subindices of the E-Government Development Index (EGDI), the Online Service Index (OSI). Also, it does not provide data for a detailed analysis on a country-level;

    3) ODIN is more focused on statistics evaluating data coverage in terms of the availability of statistical indicators in selected categories of social, economic, and environmental statistics (22 in total) and the openness of these datasets. However, it does not cover the impact and understanding of or reflection on how open data can be or are used to create value;

    4) ODMR focuses on the maturity of open data. It seems to be the most relevant benchmark since it provides detailed data and information, which can be transformed into knowledge about this topic in European countries through the years.

## 2.3 Disparities and divides in benchmarking of open data initiatives

There can be significant disparities and divides in benchmarking open data initiatives, hindering their effectiveness and impact. They are usually categorized in the context of ICT disparities and digital divides, and e-government development because open data are considered as one of the services of e-government and share some of the same resources (Lnenicka and Machova, 2022; Susha et al., 2015).

Except for the ODIN index, which is limited to analyzing the openness of statistics, all rankings refer to three main pillars: *policy*, *impact*, and *central portal*. The emphases in each ranking are distributed differently. Still, a deeper analysis reveals a far-reaching similarity in the composition of indicators, with the main difference, as a rule, being in the weights assigned to them and the way they are combined. They, in turn, can change within the same index over the years, which is also proved by recent editions of well-established and well-maintained indices compared to their previous editions.

We can recognize a number of dimensions of the ODE that have been included in the framework in recent years. They are associated with procedures that encourage user participation, cooperation, and/or collaboration to (co-)create value. Most are made available through open data portals or channels and platforms of other public sector organizations. Fundamental factors influencing what is monitored and how benchmarking frameworks are updated are sustainability and environmental challenges. It is closely related to resource centralization, green computing, and consolidated data infrastructure issues. While open data are often perceived simply as a service or resource that should be easily available to users, meet the required standards, and be free of charge, there is always a foundation of hardware, software, and human resources in place.

Disparities in the values of individual indicators used in benchmarking of open data initiatives can be contextualized in economic, geographic, legal and regulatory, technological, and merit terms, among others. For example, ODIN assesses the coverage and openness of 22 categories of statistics in 3 categories: Social (median 47), Environmental (median 48.8), and Economic and Financial Statistics (median 63.1), with the latter consistently receiving the highest overall scores. This means that in one particular subject category, the process of opening data is more advanced - a separate question is whether this is a supply effect, a demand effect, or related to, for example, the higher value of the data in that category. In the case of ODIN, most of the disparities are, however, connected to the economic context. Low- and middle-income economies are falling farther behind, with the median score in this group decreasing compared to the previous edition. At the same time, a long-term analysis spanned over seven years proves that countries from Eastern Asia have made the most progress since 2016. Data presented in the ODIN report prove that countries from the same region usually follow the same path toward openness.

The OGD underlines geographical disparities, with Europe leading the process of opening data in all categories including: health, education, employment, social security, environment, and justice. This success can be explained by implementing the EU's regulations on open data and supporting many regional initiatives. The American continent follows Europe in three categories: justice, employment, and health; Asia in two categories: education and social security, with Africa falling last in all categories.

Other axes of disproportionality between countries may result from the specific structure of composite indicators and the dimensions/pillars defined within them. This phenomenon is observable even for countries from a single region, as in the most recent ODMR'2022 report. Within the Policy dimension, several countries achieve almost perfect scores of 98-99% (Cyprus, Spain, Ireland, Poland, Italy), while others are visibly lacking – Romania (68%), Luxembourg (62%), and Malta (50%). The Portal dimension identifies three top countries: France (100%), Poland (99%), and Ireland (97%), with Malta (47%) and Slovakia (46%) closing the list. At the same time, the Quality dimension offers the smallest disparities, as most countries achieve scores from 61 to 93%, with only Malta getting 48%. Finally, the Impact Dimension shows even bigger disparities, with five countries receiving a maximum score of 100% (Cyprus, Czech Republic, Estonia, France, Ireland) and three countries ranking below 30%: Latvia (24%), Malta (18%) and Greece (12%). Generally, countries ranking well in one pillar also rank higher in other pillars, proving that this is the result of even development in different areas of building an ODE, where synergies between initiatives classified in different pillars are crucial. However, one of the most interesting cases is Greece, which achieves a somewhat satisfactory score of 77% in the portal dimension, while ranking last in the impact dimension.

## 3 Research methodology

As mentioned above, two RQs were developed to identify patterns observed in open (government) data initiatives over the years and evaluate their effects that could lead to disparities in the development and benchmarking of ODEs. The RQ1 aims to identify the patterns observed in open data initiatives over the years. To do this, we need to identify existing benchmarks, indices, and rankings of open (government) data initiatives and in what contexts these initiatives are shaped. The RQ2 deals with the effects of the identified patterns that could lead to disparities and divides in the development and benchmarking of ODEs.

To answer our research questions, we followed the methodological steps presented in Figure 1: (1) literature review to establish a knowledge base and identify contexts that have been found to shape open (government) data initiatives; (2) development of the study protocol, which content is based on the outputs of the first step, sample selection, and creation of an expert panel; (3) data collection, that is a completion of the protocol developed in the previous step by the established expert panel, and evaluation of these protocols (as one of the steps of the Delphi method); (4) identification and validation of development patterns as a result of the analysis of completed protocols in two rounds of the Delphi process, and development of recommendations based on the conducted analysis and identified patterns.

In more detail, we, first, conducted a literature review, which allowed us to identify several crucial works. Our analysis was the basis for the selection of a sample of information artifacts as it allowed us to identify several relevant contexts, which can be divided in two groups: (1) directly related to open (government) data - including open data indices and rankings, national OGD strategies, and other documents or benchmarks related to OGD at a more regional or national level), (2) those affecting and/or shaping open (government) data initiatives - e.g., e-government, digital readiness, emerging topics and innovations, sustainable development-oriented movements. These, however, may vary from one country to another. These contexts set the general structure of the protocol, which was further refined in step 2.

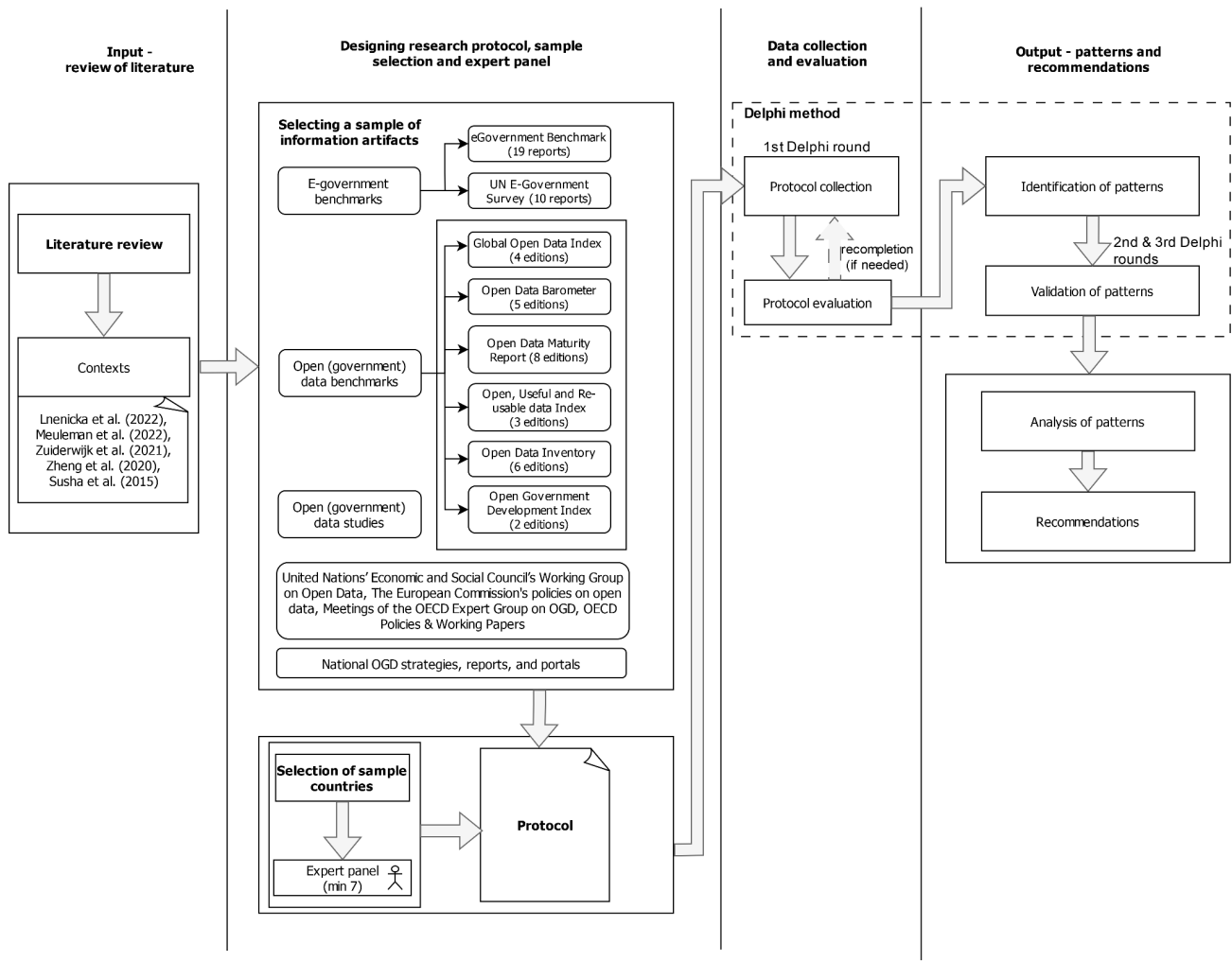

*Figure 1.* Methodological steps of the study

The refinement process included analysis of relevant resources: e-government benchmarks (UN E-Government Survey 2022 - 10 reports between 2003 and 2020, eGovernment Benchmark - 19 reports between 2001 and 2021), open (government) data benchmarks (GODI - 4 editions, ODB - 5 editions, ODMR - 8 editions, OURData - 3 editions, ODIN - 6 editions and OGDI - 2 editions) and other relevant sources. Finally, we have updated the results published in Lnenicka et al. (2022) in the case of indices, for which new editions have been released since then. This update was necessary for the comparative analysis of the sampled countries.

Prior research has determined that organizational change (Jacobs et al., 2013) and technology implementation, especially open data initiatives, is content-dependent (Sayogo and Pardo, 2012; Zuiderwijk et al., 2015). Therefore, considering the complexity and the variety of the identified contexts, an in-depth analysis of countries is required, including e-government, open government, open (government) data, as well as factors that may affect the above, but not necessarily directly related to them (i.e., cultural, political, economic and/or historical specificity), and potential effects of having reports and/or indices with more limited scope and/or regional coverage (country, region). Such a wide range of knowledge and depth of research implies not only the determination of the patterns we are looking for but also, possibly, the limitation of the set of these factors to be studied in the future when replicating or reproducing the study or maintaining its results.

For this reason, we have adapted the approach Breaugh et al. (2023) used to select representative countries. A cross-country case study methodology has been developed to dive deeper into individual services (Yin, 2018; Mergel et al., 2019, p. 12) without losing sight of the bigger picture (Lijphart, 1971). Exploratory case studies are especially useful when there is a need to develop new hypotheses and propositions, particularly when the issue of study is contemporary with limited empirical information available (Chopard and Przybylski, 2021; Eisenhardt, 1989; Yin, 2018).

Eight countries were selected based on several criteria. The primary criterion was the country's presence in the analyzed reports. To diversify the list, we selected competitive and less competitive countries. This allowed us to avoid having only the most competitive countries according to these reports and/or indices, which would prevent us from considering the pitfalls faced by less competitive countries. Contrary, having only the lowest performing countries would mean not being able to consider best practices. Moreover, we selected countries that are always covered by the selected indices and reports, as well as those that tend to be represented only in some of them. The first case allows to track their progress and base their agenda for their development on published documents, as well as benchmarking results over the years and/or compared to other countries set as benchmarks. The second case allows us to understand whether coverage by these indices and reports affects the state of affairs. This choice is based on the results of our literature review covered in the previous section.

To this end, we collected data for all editions of GODI, ODB, OURdata Index, ODIN, ODMR, and OGDI, including the list of countries they cover and countries' results in these indices. Then we analyzed how often these indices covered a country, thus identifying those for which only a few values are missing (a few editions did not cover it) or a prevailing number of values is missing with reference to European countries only (see Table 2). Then, the results of the countries shown in these rankings were analyzed. Additionally, we ensured that the countries we selected reflected different administrative traditions, geographical areas, and unitary and federal states (in line with Kuhlmann and Wollmann (2019)). Based on the above steps, eight countries made up our sample.

*Table 2.* Representativeness of countries in rankings

| **Representativeness level** | **List of countries** |
|---|---|
| Well represented (max 2 out of 28 missing values) | France, Germany, **Italy** |
| Well represented (3 to 6 out of 28 missing values) | **Austria**, **Belgium**, Denmark, Finland, Greece, Netherlands, Norway, Portugal, **Sweden**, Switzerland, United Kingdom, **Czech Republic,** Ireland, Spain |
| Moderately represented (7 to 16 out of 28 missing values) | **Poland,** Slovakia, Estonia, Bulgaria, **Latvia,** Slovenia, Hungary, Croatia, Iceland, Romania |
| Poorly represented (more than 16 missing values) | Israel, Lithuania, Ukraine, Cyprus, Luxembourg, Malta, **Serbia,** Albania, Bosnia and Herzegovina, Liechtenstein, Kosovo (under UN1244 resolution), Belarus |

Bold style is applied to the country selected from the respective category to constitute the sample

We then established an *expert panel* representing each country in the sample. By the term "expert" we mean a person possessing both in-depth knowledge of the subject and the context associated with the specifics of a particular country that might affect the results. It could be the country of origin of the expert or the country where the expert is employed or involved in open data initiatives. This means that the person must have at least a master's degree in the field related to at least some OGD aspects, familiar with others at the same time (e.g., business and management, political sciences, law, computer sciences, etc.), with at least five years research and practical experience related to OGD projects and/or OGD initiatives in public administration of the country in question.

To answer the RQs, we then applied the 10 steps of the Delphi method as proposed by Linstone (1985). The first two steps are covered by the selection of experts and formation of the expert panel. Our panel size of eight experts is compliant with Linstone (1985), according to which a suitable minimum panel size is seven with accuracy deteriorating rapidly with smaller sizes and improving more slowly with a larger number (Mullen, 2003). The next two steps consisted of the development of the protocol and testing the proper wording (vagueness, redundancies etc.). This step was conducted by two authors. One of them previously acted as a facilitator of the Delphi process too.

In the next step, i.e., the first round of data collection for each country using the protocol (Annex 1), we aimed to get a clearer view of (1) how national open data initiatives performed over the years in selected countries, i.e., their results and trends in the respective indices and sub-indices; (2) what open data indices and rankings (if any) are taken into account by selected countries developing their open data initiatives; (3) what are the patterns observed in the development and benchmarking of countries in defined contexts over the years. The contexts were (1) e-government, (2) OGD, (3) open data indices and rankings, and (4) other relevant resources, including but not limited to reports, reviews, indices, and rankings of national and regional importance, dealing with developments and setting the agenda/future steps for developing the initiative. This step was performed between December 2022 and January 2023. The next step was the analysis of the first round of responses, including whether and how the various benchmarks contributed to understanding the state of open (government) data initiatives over the years. The patterns for each context were determined by analyzing the results of eight protocols. For this study, we define the pattern as "*an activity or milestone identified at least in one country and has a positive or negative implication for the benchmarking and development of open data initiatives in the country*." This step was done in January 2023.

The second round consisted of validating the original list of determined patterns, i.e., whether the pattern occurs/is observed in a particular country and evaluating their potential impact on developing and benchmarking the country's ODE. This was done by all eight experts, with each expert also being asked to add new patterns relevant to their country or clarify, reformulate, or merge a pattern with another, or split a pattern into multiple patterns. In the third round, new and updated patterns and the effects that could lead to disparities were validated again, and the agreement was reached on a final list of developed patterns. This step was performed in February 2023. The response rate was 100% for all three rounds of the Delphi process, there were no dropouts of experts. The last step of the Delphi process included the preparation of the list of patterns for the cluster analysis, i.e., their coding and formatting.

The last step of our approach, the analysis of patterns, involved clustering patterns based on the similarities of patterns observed for each context. With this, we aimed to understand whether it is possible to determine clusters based on prevailing common patterns. This would allow us to identify strengths and weaknesses that may be recognized as best practices or lead to disparities in benchmarking open data initiatives. Since our study is exploratory in nature, and we use a relatively small sample of countries, we obtained findings that can later be used to validate them on a larger sample by identifying these clusters. Finally, based on the findings, two sets of recommendations were derived for further actions to promote the development of the ODEs. The first set of recommendations is directly related to the three primary contexts associated with OGD and open government, and therefore targeted at public administration. The second set forms high-level recommendations that were derived from the patterns identified for *context D*. They aim to increase the sustainability and resilience of OGD initiatives and ecosystems. Both sets of descriptions are based on best practices that were observed in selected countries, i.e., these practices are considered critical for success.

In other words, our study follows the approach used in Styrin et al. (2017) - our analysis is based on a study of indices and benchmarks of open data initiatives, local and regional documents of countries under review, such as OGD strategies, strategies and action plans for national development, future projects, and national/global trends, open data portals in every selected country. They were supplemented by other sources of information that proved relevant in the context of a particular country and through personal conversations between the authors and relevant officials of government organizations. Therefore, it can be said that in identifying and confirming the identified patterns, similarly to Styrin et al. (2017), we use a comparative approach (Rose and Mackenzie, 1991), which involves using a set of common concepts for a group of selected countries to analyze similarities and disparities within this group. Comparative study examines group phenomena that vary across countries using ideas or shared frames of reference. Our searches for relevant documents were based on a set of concepts, benchmarks, or criteria often used in studies of e-government and OGD maturity and empirical analysis of ODEs.

## 4   Case study - collected data on the contexts shaping open data initiatives

Table 3 presents the indicators about the selected countries provided by Eurostat (data from 2021 or 2022). It should be noted that these characteristics can affect whether a country is included in the benchmark. Some benchmarking initiatives cover only developed countries or countries that have adopted open data principles, such as the Open Data Charter or Open Government Partnership (OGP).

*Table 3.* General data about sample countries

| Country | Population | Population density (per km$^2$) | Area (km$^2$) | GDP nominal (EUR mil) | GDP per capita (EUR) | EU Member / since when |
|---|---|---|---|---|---|---|
| Austria | 9 090 868 | 108 | 83 883 | 406 148.7 | 45 370 | 1995 |
| Belgium | 11 584 008 | 377 | 30 688 | 502 311.6 | 43 330 | 1958 |
| Czech Republic | 10 525 739 | 139 | 78 871 | 238 249.5 | 22 270 | 2004 |
| Italy | 58 983 122 | 195 | 302 068 | 1 782 050.4 | 30 140 | 1958 |
| Latvia | 1 834 588 | 30 | 62 200 | 33 695.9 | 17 890 | 2004 |
| Poland | 37 990 000 | 121 | 312 696 | 574 771.8 | 15 060 | 2004 |
| Serbia | 6 797 105 | 93 | 88 499 | 53 329.3 | 7 800 | candidate |
| Sweden | 10 512 820 | 26 | 447 425 | 537 085.0 | 51 560 | 1995 |

## 4.1 E-government context

Figure 2 shows the progress of the EGDI in the sample countries between 2003 and 2022. As stated above, the state of e-government development, and respective digital public services and projects, represents the external pressure that affects the development of the ODE. Although the methodology of the EGDI slightly changed over the years, all the countries constantly improved their results. We can argue that the current state of e-government in sample countries is so developed that it could only be enhanced by new ways of providing services, e.g., in the metaverse platform.

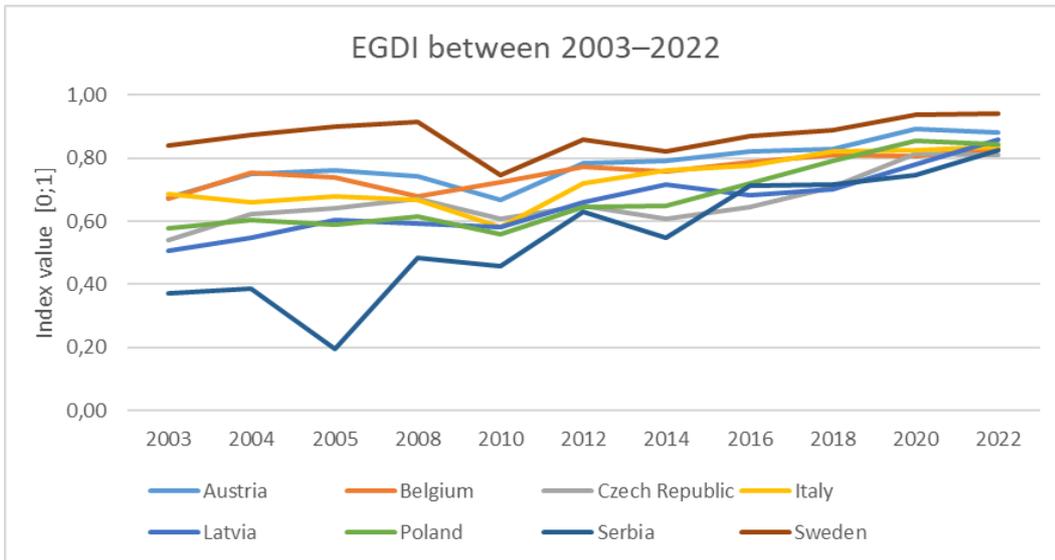

*Figure 2.* Progress of the EGDI in sample countries

Since the provision of open (government) data is often considered a digital public service, it is worth taking a closer look at the OSI. It is one of the sub-indices of the EGDI, which evaluates the scope and quality of online services and can provide us with data on the progress of all services in the country. The quality of services in sample countries improved over the years (Figure 3), but there are still areas for improvement, especially in Belgium and the Czech Republic.

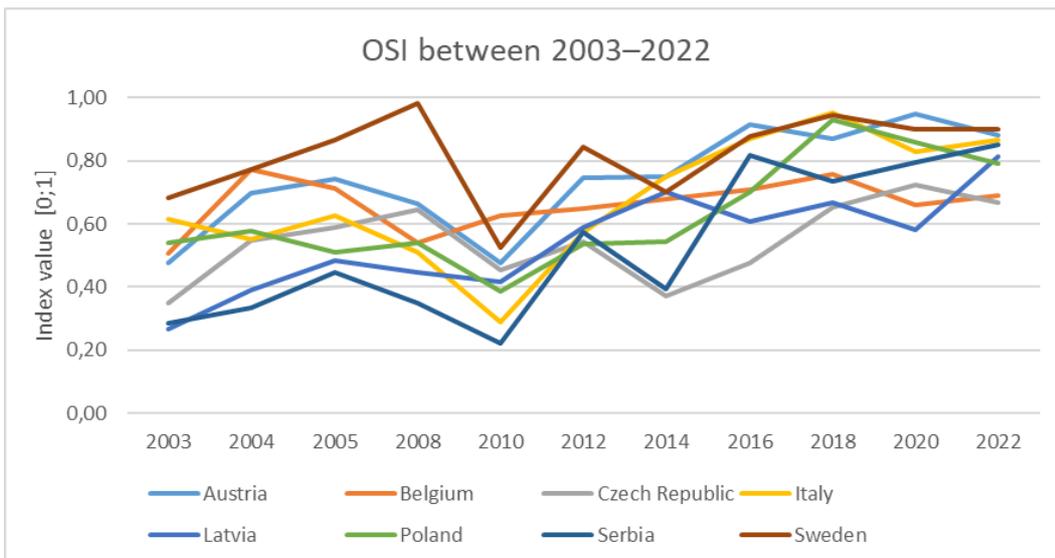

*Figure 3.* Progress of the OSI in sample countries

Because the development status of telecommunication infrastructure is key for data publishing and sharing, Figure 4 presents the progress of the next sub-index of the EGDI, the Telecommunication

Infrastructure Index (TII), between 2003–2022. Compared to 2003, we can state that the quality of infrastructure has improved significantly in all countries, and the access of all stakeholders to services is provided.

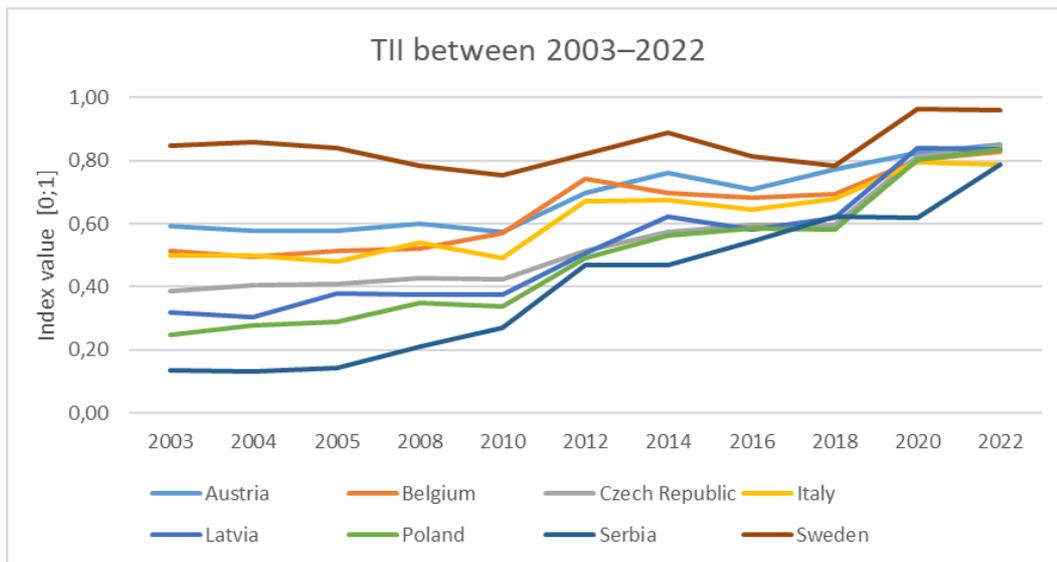

*Figure 4.* Progress of the TII in sample countries

### 4.2 Open (government) data context

In this section, we focus on the OGD strategy and the link between OGD efforts and open data benchmarks in the country. To this end, we aim to understand (1) whether the country had an OGD strategy, (2) when the strategy was first published, and (3) whether an active strategy is in use. In addition, we also investigated (4) when open (government) data first emerged in IT/smart strategies, i.e., strategies other than the OGD strategy, mainly for those countries that do not have an OGD strategy or have not had one for some time, (5) whether the results of OGD assessment by benchmark (reports) are used to set the agenda and corrective action, and (6) which benchmark is used for this purpose, serving as the most important source of information in this area. We also analyzed how many open data indices and reports evaluate or measure open data efforts in the country. The results of this investigation are summarized in Table 4.

As part of our analysis, we found that the lack of an OGD strategy in a country is usually due to the fact that (1) the topic of open data is included in national legislation and laws. For the EU's context, it is set by the relevant directives, i.e., the directive on open data and the re-use of public sector information and/or (2) the topic is included in other strategies, e.g., ICT and digital strategies and/or OGP action plans. We found that the OGP, which was launched in 2011, and the action plans that are developed by its members, usually for a two- or three- year period, play an important role in the first appearance of the open data topic in national strategies and for some countries, these action plans act as OGD strategies. Most countries also provide guidelines and handbooks for governments and civil society on data opening and reuse, e.g., the Austrian guide Open Data Governance – Towards a Data-Driven Organization or the Serbian guide to open data. Free online courses on open data are also very often provided too.

In terms of the indices and reports, from the results/outputs used to set the agenda and corrective actions for OGD efforts, only a few sample countries published their respective analyses. One of them is the Czech Republic, which has published annual reports on the status of open data publication every year since 2017. Other countries usually rely on a series of reports published by the EU, i.e., ODMR, which are very detailed and provide sufficient information for most countries, so they do not produce nor publish other reports. However, we should mention that while many OGD strategies refer to an OGD index or benchmark (mostly ODMR), usually it is rather a mention and not a real basis for defining corrective actions (i.e., is the case for those where "yes" appears in Table 4). Finally, there are only a few active national OGD strategies in 2022 because most countries prefer strategies with a wider scope, i.e., also covering big data, Artificial Intelligence (AI), etc., as in the case of Belgium or Sweden.

*Table 4.* Overview of OGD strategies' related information

| Country | Austria | Belgium | Czech Republic | Italy | Latvia | Poland | Serbia | Sweden |
|---|---|---|---|---|---|---|---|---|
| Is there any OGD strategy in the country? | NO | YES | NO | NO | YES | YES | NO | NO |
| When was the first OGD strategy published? | N/A | 2015 | N/A | N/A | 2019 | 2016 | N/A | N/A |
| Is there a valid OGD strategy in 2022? | NO | NO | NO | NO | YES | YES | NO | NO |
| When did open (government) data first appear in IT / smart strategies? | 2012 | N/A | 2012 | 2011 | 2013 | N/A | 2018 | 2012 |
| Is the result of an OGD effort as assessed by benchmark (reports) used to set the agenda and corrective action? | YES | N/A | YES | N/A | YES | YES | YES | N/A |
| What benchmark is used for this purpose, i.e., is the most important source of information in this area? | ODMR | ODMR | ODMR | ODMR | ODMR | ODMR | GODI | ODMR |

GODI - Global Open Data Index, ODMR - Open Data Maturity Report (also European Open Data Maturity Assessment)

We then focused on the OGD portals in the sample countries. We wanted to find when the first official national OGD portal was launched, whether there were any prior efforts (portals), and how the portal has evolved over the years, etc. The results of this analysis are summarized in Table 5. The investigation took place in March 2023.

We found that several waves are seen here. The first wave of launching national portals began after 2011 and could be related to the OGP. The second wave can be described as a response to the ODMR, which was first published in 2015, and the availability and quality of an open data portal are one of the indicators ODMR covers. However, there is no fixed pattern, and most countries launched portals at their own pace. It was also found that in many countries, an unofficial portal existed before the official one was launched. Most of these were CKAN-based portals, which is probably because this data management system is open-source and easy to deploy and manage. Datasets that initially did not follow the open data were gathered from external websites of public agencies and public sector institutions and published on these unofficial portals. Several portals were also launched in some cities, regions, and federal states, but most disappeared or were later merged into a national portal.

Regarding the number of datasets, the countries in the sample are actively publishing data. However, if we analyze categories, providers, or the presence of high-value datasets, we observe that a higher number does not correlate with greater openness. For example, on the Czech national open data portal, more than 95% of the datasets were published by the Czech Office for Surveying, Mapping, and Cadastre. In addition, it categorizes datasets into 197 themes/topics, which is also very confusing. On the other hand, the Polish portal allows the filtering of high-value datasets. Generally, most portals provide features for working with datasets, but their quality and usability need to be studied, where we analyzed only their availability. Half the portals do not provide a section on reuses (showcases/use cases) or applications built using open data.

*Table 5.* Overview of OGD portals' related information

| Country | Austria | Belgium | Czech Republic | Italy | Latvia | Poland | Serbia | Sweden |
|---|---|---|---|---|---|---|---|---|
| Launch of the first official national OGD portal. | 2012 | 2015 | 2018 | 2011 | 2017 | 2014 | 2018 | 2012 |
| Was there any unofficial portal before the official one was launched? | YES | YES | YES | NO | NO | YES | YES | NO |
| Number of datasets on the portal. | 42796 | 16089 | 142383 | 56608 | 725 | 30755 | 2168 | 8102 |
| Number of categories / themes / topics. | 17 | 14 | 197 | 13 | 14 | 14 | 9 | 13 |
| Number of organizations / data providers. | 2395 | 107 | 287 | 928 | 96 | 286 | 111 | 214 |
| Number of reuses / applications. | 708 | 80 | N/A | N/A | N/A | 70 | 40 | N/A |

### 4.3 Open data indices and rankings context

Table 6 presents an overview of the countries in the context of selected indices and reports, providing respective ranks and values. In total, we considered 27 editions of selected open data indices and reports (data for the ODMR from 2015 are not available) (data on all countries covered by studied indices are available on Zenodo[1]). Because the number of countries covered by indices differed over the years, we added the quartiles for each rank.

*Table 6:* Overview of ranks and values for open data indices in sample countries

| Index | | Austria | Belgium | Czech Republic | Italy | Latvia | Poland | Serbia | Sweden |
|---|---|---|---|---|---|---|---|---|---|
| GODI 2013 | Rank [out of 60] | 23 (Q2) | 56 (Q4) | 29 (Q2) | 21 (Q2) | N/A | 36 (Q3) | 31 (Q3) | 8 (Q1) |
| | Value [0,100] | 51 | 27 | 45 | 52 | N/A | 42 | 44 | 67 |

---
[1] https://zenodo.org/doi/10.5281/zenodo.10231024

| | | | | | | | | |
|---|---|---|---|---|---|---|---|---|
| GODI 2014 | Rank [out of 97] | 23 (Q1) | 53 (Q3) | 13 (Q1) | 25 (Q2) | 34 (Q2) | 48 (Q2) | 48 (Q2) | 13 (Q1) |
| | Value [0,100] | 59 | 39 | 66 | 55 | 51 | 42 | 42 | 66 |
| GODI 2015 | Rank [out of 122] | 23 (Q1) | 35 (Q2) | 21 (Q1) | 17 (Q1) | 31 (Q1) | N/A | N/A | 27 (Q1) |
| | Value [0,100] | 50 | 43 | 52 | 55 | 46 | N/A | N/A | 48 |
| GODI 2016 | Rank [out of 94] | 28 (Q2) | 22 (Q1) | 27 (Q2) | 32 (Q2) | 14 (Q1) | 28 (Q2) | 41 (Q2) | 21 (Q1) |
| | Value [0,100] | 49 | 52 | 50 | 47 | 64 | 49 | 41 | 53 |
| ODB 2013 | Rank [out of 77] | 18 (Q1) | 31 (Q2) | 22 (Q2) | 20 (Q2) | N/A | N/A | N/A | 3 (Q1) |
| | Value [0,100] | 46.0 | 34.8 | 43.2 | 45.3 | N/A | N/A | N/A | 85.8 |
| ODB 2014 | Rank [out of 86] | 15 (Q1) | 27 (Q2) | 17 (Q1) | 22 (Q2) | N/A | 35 (Q2) | N/A | 3 (Q1) |
| | Value [0,100] | 58.52 | 47.29 | 58.07 | 50.58 | N/A | 36.99 | N/A | 83.7 |
| ODB 2015 | Rank [out of 92] | 13 (Q1) | 22 (Q1) | 26 (Q2) | 21 (Q1) | N/A | 32 (Q2) | N/A | 9 (Q1) |
| | Value [0,100] | 64.18 | 52.62 | 49.15 | 53.78 | N/A | 39.95 | N/A | 69.26 |
| ODB 2016 | Rank [out of 115] | 14 (Q1) | 29 (Q2) | 31 (Q2) | 20 (Q1) | 53 (Q2) | 46 (Q2) | 65 (Q3) | 14 (Q1) |
| | Value [0,100] | 70.22 | 45.28 | 44.44 | 55.93 | 27.89 | 33.95 | 22.77 | 69.84 |
| ODB 2017 | Rank [out of 30] | N/A | N/A | N/A | 14 (Q2) | N/A | N/A | N/A | N/A |
| | Value [0,100] | N/A | N/A | N/A | 50 | N/A | N/A | N/A | N/A |
| OURdata Index 2015 | Rank [out of 30] | 14 (Q2) | 18 (Q3) | N/A | 24 (Q4) | N/A | 29 (Q4) | N/A | 28 (Q4) |
| | Value [0,1] | 0.62 | 0.54 | N/A | 0.39 | N/A | 0.13 | N/A | 0.24 |
| OURdata Index 2017 | Rank [out of 34] | 9 (Q2) | 22 (Q3) | 23 (Q3) | 19 (Q3) | 33 (Q4) | 20 (Q3) | N/A | 30 (Q4) |
| | Value [0,1] | 0.68 | 0.46 | 0.45 | 0.52 | 0.19 | 0.48 | N/A | 0.31 |
| OURdata Index 2019 | Rank [out of 33] | 12 (Q2) | 20 (Q3) | 17 (Q3) | 18 (Q3) | 22 (Q3) | 15 (Q2) | N/A | 32 (Q4) |
| | Value [0,1] | 0.65 | 0.58 | 0.61 | 0.60 | 0.54 | 0.63 | N/A | 0.38 |
| ODIN 2015 | Rank [out of 125] | N/A | N/A | N/A | N/A | N/A | N/A | 11 (Q1) | N/A |
| | Value [0,100] | N/A | N/A | N/A | N/A | N/A | N/A | 47.6 | N/A |
| ODIN 2016 | Rank [out of 173] | 19 (Q1) | 44 (Q2) | 2 (Q1) | 11 (Q1) | 12 (Q1) | 4 (Q1) | 39 (Q1) | 1 (Q1) |
| | Value [0,100] | 63.4 | 51.8 | 79.1 | 71.5 | 71.4 | 77.5 | 53.2 | 81.0 |
| ODIN 2017 | Rank [out of 180] | 35 (Q1) | 56 (Q2) | 16 (Q1) | 23 (Q1) | 18 (Q1) | 4 (Q1) | 94 (Q3) | 3 (Q1) |
| | Value [0,100] | 56.1 | 49.1 | 67.1 | 62.5 | 65.6 | 75.4 | 37.1 | 77.3 |
| ODIN 2018 | Rank [out of 178] | 24 (Q1) | 73 (Q2) | 20 (Q1) | 30 (Q1) | 34 (Q1) | 4 (Q1) | 76 (Q2) | 9 (Q1) |
| | Value [0,100] | 67.7 | 49.0 | 69.6 | 64.6 | 62.0 | 82.5 | 48.1 | 78.1 |
| ODIN 2020 | Rank [out of 187] | 30 (Q1) | 113 (Q3) | 27 (Q1) | 37 (Q1) | 60 (Q2) | 2 (Q1) | 46 (Q1) | 5 (Q1) |
| | Value [0,100] | 68.7 | 45.5 | 69.8 | 65.9 | 58.2 | 85.3 | 62.6 | 83.9 |
| ODIN 2022 | Rank [out of 192] | 55 (Q2) | 68 (Q2) | 33 (Q1) | 35 (Q1) | 16 (Q1) | 4 (Q1) | 31 (Q1) | 10 (Q1) |
| | Value [0,100] | 60.9 | 57.0 | 68.3 | 67.8 | 75.4 | 85.7 | 68.7 | 80.0 |
| ODMR 2016 | Rank [out of 31] | 5 (Q1) | 23 (Q3) | 18 (Q3) | 20 (Q3) | 30 (Q4) | 17 (Q3) | N/A | 24 (Q4) |
| | Value [0,1] | 0.78 | 0.48 | 0.55 | 0.52 | 0.15 | 0.56 | N/A | 0.44 |
| ODMR 2017 | Rank [out of 32] | 13 (Q2) | 20 (Q3) | 21 (Q3) | 8 (Q1) | 19 (Q3) | 23 (Q3) | N/A | 22 (Q3) |
| | Value [0,1] | 0.77 | 0.68 | 0.68 | 0.81 | 0.68 | 0.62 | N/A | 0.65 |
| ODMR 2018 | Rank [out of 31] | 16 (Q3) | 15 (Q2) | 21 (Q3) | 4 (Q1) | 12 (Q2) | 13 (Q2) | N/A | 23 (Q3) |
| | Value [0,1] | 0.65 | 0.65 | 0.62 | 0.80 | 0.66 | 0.66 | N/A | 0.52 |
| ODMR 2019 | Rank [out of 32] | 16 (Q2) | 17 (Q3) | 19 (Q3) | 8 (Q1) | 11 (Q2) | 7 (Q1) | N/A | 24 (Q3) |
| | Value [0,1] | 0.66 | 0.65 | 0.64 | 0.77 | 0.75 | 0.78 | N/A | 0.55 |
| ODMR 2020 | Rank [out of 35] | 7 (Q1) | 24 (Q3) | 21 (Q3) | 9 (Q2) | 19 (Q3) | 6 (Q1) | N/A | 16 (Q2) |
| | Value [0,1] | 0.90 | 0.62 | 0.72 | 0.87 | 0.80 | 0.90 | N/A | 0.84 |
| ODMR 2021 | Rank [out of 34] | 7 (Q1) | 30 (Q4) | 23 (Q3) | 8 (Q1) | 21 (Q3) | 4 (Q1) | N/A | 17 (Q2) |
| | Value [0,1] | 0.92 | 0.55 | 0.74 | 0.92 | 0.77 | 0.95 | N/A | 0.84 |
| ODMR 2022 | Rank [out of 35] | 17 (Q2) | 25 (Q3) | 12 (Q2) | 8 (Q1) | 30 (Q4) | 3 (Q1) | 27 (Q4) | 18 (Q3) |
| | Value [0,1] | 0.79 | 0.69 | 0.88 | 0.91 | 0.57 | 0.95 | 0.66 | 0.78 |
| OGDI 2020 | Rank [out of 191] | 1 (Q1) | 44 (Q1) | 1 (Q1) | 1 (Q1) | 57 (Q2) | 44 (Q1) | 65 (Q2) | 1 (Q1) |
| | Value [0,1] | 1.00 | 0.93 | 1.00 | 1.00 | 0.86 | 0.93 | 0.85 | 1.00 |
| OGDI | Rank [out of | 26 (Q1) | 74 (Q2) | 19 (Q1) | 12 (Q1) | 19 (Q1) | 69 (Q2) | 26 (Q1) | 1 (Q1) |

| 2022 | 193] | | | | | | | | |
|---|---|---|---|---|---|---|---|---|---|
| | Value [0,1] | 0.94 | 0.73 | 0.97 | 0.99 | 0.97 | 0.76 | 0.94 | 1.00 |

The GODI covered all countries in the sample at least three times and measured selected data categories' legal and technical openness. Since the number of categories increased from 10 in 2013 to 15 in 2016, this resulted in changes in weights, and the comparability of the final scores over the years is limited. Some sectors and related categories included in the set were open, while others were closed. The reasons for this are related to the different priorities of countries in opening data from the respective sectors, legal restrictions, and lack of cooperation of some authorities in this process.

The ODMR and ODIN can be considered the most detailed indices for benchmarking open data initiatives. We can argue that they are used by the holders of the open data initiative when reflecting on the current state of development of the open data initiative and when planning an additional set of activities, including corrective actions. The ODIN index places Sweden in the top tier with very high scores for openness and lower for coverage. The ODMR has increased the rank for Sweden from the last quartile towards average performance. None of the indicators stand out. These two indices also positively assess the efforts of Serbia, which showed an orientation towards a higher position in the indices and rankings. The positive dynamics observed in recent years may be due to various factors. One is the implementation of a new e-government law launched in 2018, as well as its improved commitment to open data. To this end, the ODIN index shows an upward trend in Serbia's ranking in its recent editions. In addition, the country's efforts to establish an OGD portal and promote new related strategies contributed to improved rankings, including relatively positive results in ODMR, as well as various studies that have been carried out in the country, focusing on assessing the impact of open data. Alongside these efforts, Serbia intends to explore ways to involve the private sector in the open data initiative. There are also attempts to support public sector organizations and higher education institutions that are involved in projects related to the reuse of open data and the promotion of open data. To facilitate the reuse of open data by citizens, Serbia has established monitoring processes through its national open data portal. In addition, they have either initiated or planned activities aimed at encouraging government organizations to track the reuse of their own published data.

Surprisingly, it can be noticed the stagnation of the development of OGD in Austria. In particular, the missing implementation of the Freedom of Information Act affected its ranking in ODMR, decreasing to the follower category in 2022 (ranked 17), while in previous years, it was ranked as fast-tracker (5-16). The decrease in the completeness and impact dimensions of the report is noticeable and again suggests stagnating open data initiatives in Austria. Similar, results in ODMR for Latvia are uneven, with the increase in the value and rank with the launch of the national OGD portal (2017*) and some interest from stakeholders and government, including the development of the national open data strategy (2021 – very late), but unfortunately decreased in ranking in recent years, especially the last one (2022). Although, at the same time, it should be noted that the value itself was increasing in some of these years (when the rank decreased), i.e., the initiative developed but not as fast as others. In the recent edition, Latvia lost a lot in both rank and values and reached the lowest result ever.

Furthermore, ODIN shows a negative trend for Italy which lost points and position over the years, ranking at 37 in 2020. Italy lags behind the coverage with regard to some sectors, including the built environment (with zero coverage), agriculture and land use, poverty and income, food security and nutrition, health outcomes, and health facilities. The ODIN confirms the good results regarding coverage and openness of GODI and ODB regarding census data (population and vital statistics). For Poland, ODIN (and other individual reports) only prove that (depending on the methodology adopted) the development of the open data system allows it to maintain more or less the same position in subsequent reports. It is noticeable that only Italy and the Czech Republic maintain more or less the same position in comparison with each index and ranking, while the other countries vary in values.

The different nature of each index and ranking described in previous sections clearly indicate the discrepancy of rankings and the variety in such results. Another important fact that can explain this situation is that there is maybe a pressure to look good in comparison to others, which can make the data collected unreliable. In addition, one of the main problems with most of those reports is that they tend to be based at least partially on self-evaluation reports. Differences in the assessments suggest that a comprehensive approach that looks at contents and coverage, as well as policy/governance and infrastructure, is still lacking. Lack of transparency and consistency across different levels of government, which makes it difficult for citizens and organizations to find and access the data they need – the non-availability of certain data also influences the rankings in open data indices. The question is if citizens are active/activated in terms of the democratic processes and participation in decision-making processes and interested in open data strategies, i.e., openness and transparency topics.

## 5  Case study - summary and findings

### 5.1  Patterns, their occurrence, and their effects

Three rounds of the Delphi method were performed to determine the final list of patterns. One hundred two (102) patterns for all examined contexts were identified for at least one country in the first round. In the second round, the occurrence and effects/impact on the development and benchmarking of the country's ODE with respect to each pattern were evaluated for each country. Each pattern was subject for clarification, refinement, reformulation, merge with another, or split into multiple patterns. In this round, thirteen (13) patterns were reformulated, and eight (8) patterns were removed from the list. These changes were validated and approved in the third round. The total number of 94 patterns in four groups, i.e., contexts, was obtained. Twenty patterns (20) for context A refer to e-government, eighteen (18) for context B - OGD, thirty-one (31) for context C - open data indices and rankings, and twenty-five (25) for context D - other relevant resources.

The occurrence of the pattern was evaluated as a Boolean value, i.e., "YES" - the pattern occurs in the country under consideration or "NO" - the pattern does not occur in the country under consideration (1/0). The effect on the development of the country was evaluated using a 5-point Likert scale (no effect = 0, limited effect = 1, moderate/average effect = 2, significant effect = 3, extreme effect = 4). Overall, 28 patterns were found in all countries, 13 - in 7 countries, 11 - in 6 countries, and 6 - in only one country. A list of the 20 patterns with the highest effect is provided in Table 7, while a list of all patterns identified during this study is available at Zenodo[2] (with the final set of patterns in Annex 2).

More than half of the most important patterns (see Table 7) result from the e-government context. We can argue that the development of e-government, especially after 2000, and the way public services were implemented both within the internal processes of the public sector and in relation to citizens and businesses created the basis for open data. This is mainly about the data infrastructure and related processes, i.e., how public sector agencies and institutions are able to identify, pre-process, and publish their data on data portals. Benchmarking, by its very nature, is based on weaknesses and strengths in a particular area. This is to say, if the foundations were already laid in the era of e-government development, then open data initiatives were built on what existed and what needed to be improved. The ways to benchmark open data initiatives in countries have also been affected and influenced by the international and supranational environment, i.e., the EU, and what actions must be taken and what approaches must be implemented in national laws. The results also showed that not only the technological aspects and data infrastructures are important, but also the skills and motivations of all stakeholders are critical to any benchmarking efforts and improvements in development.

---
[2] https://zenodo.org/doi/10.5281/zenodo.10231024

*Table 7.* A list of 20 patterns that affected benchmarking of open data initiatives most

| Context | Pattern | Average | Median | Standard deviation |
|---|---|---|---|---|
| A | The start of building public sector information systems and base registers to enable efficient flow of information and data between public sector agencies and institutions. | 3.25 | 4.00 | 1.16 |
| B | The PSI Directives and the Open Data Directive by the EU are implemented into national law, usually in the context of free access to information rights. | 3.25 | 3.50 | 0.89 |
| A | The start of digital identity (eID) issuance and availability of digital public services that can be used in this way. | 3.13 | 3.00 | 0.64 |
| A | Continuous improvement of a centralized e-government citizens' portal. | 2.88 | 3.00 | 1.36 |
| A | Increasing interoperability of services. | 2.75 | 3.00 | 0.71 |
| A | Existence of a centralized e-government citizens' portal. | 2.75 | 3.00 | 1.28 |
| C | Digital skills are lacking for public officials. | 2.75 | 3.00 | 1.39 |
| A | Launch of the public administration portal and portals of public sector agencies and institutions with relevant and up-to-date information and life events that help citizens and businesses get the necessary information online. | 2.63 | 3.00 | 0.52 |
| A | The start of prioritization of security, reliability, and related policies for digital public services such as authentication, authorization, e-signatures etc. | 2.63 | 3.00 | 0.52 |
| C | Stakeholders – business and citizens are often either unaware of the existence of an OGD (portal), or unaware of or critical of the benefits of an OGD closed ecosystem. | 2.63 | 3.00 | 1.41 |
| A | The availability of mobile apps and access to digital public services from mobile phones (in general), including the usability and friendliness of these apps, resulted from the penetration of mobile | 2.50 | 2.50 | 0.53 |
| | phones among citizens and businesses. | | | |
| C | Digital skills and open data skills in particular are lacking for citizens. | 2.50 | 2.50 | 1.31 |
| D | The portal is reviewed and improved regularly. | 2.50 | 3.00 | 1.20 |
| A | The start of building telecommunications infrastructure and networks enabling access to the Internet as well as digital public services for all stakeholders. | 2.38 | 3.00 | 1.06 |
| A | Ensuring security of operations in the public sector, new and improved tools for authorization and authentication of citizens. | 2.38 | 3.00 | 1.06 |
| C | The datasets are accompanied with the metadata, described, and updated regularly, but the level of openness of the datasets is low (i.e., 1 to 3 stars according to a 5-star scheme). | 2.38 | 2.50 | 1.19 |
| C | The national open data portal provides support and guidelines for data reuse, but monitoring and ensuring the use of open data is often beyond the personnel and financial capacities of the country as well as regions, cities, and municipalities. | 2.38 | 3.00 | 1.30 |
| A | Increasing the availability of mobile apps provided by the public sector and the use of these apps by citizens and businesses to communicate and exchange information and data with public sector agencies and institutions. | 2.25 | 2.50 | 1.28 |
| B | The OGD as a topic is included in the national digital strategy and/or strategic documents, Action Plan dealing with digital technologies and their use – usually updated every few years. | 2.25 | 2.00 | 1.04 |
| D | The open data available on the national portal is accompanied by licensing information. | 2.25 | 2.50 | 1.04 |

{A,B,C,D} stands for the context, where A - e-government, B - OGD, C - open data indices and rankings, D - other relevant resources) (see the list of patterns in Annex 2)

We then clustered all patterns with respect to their occurrence and effects (impacts), which we determined at the previous stage, to aggregate them based on their similarities and identify groups that would allow us to identify strengths and weaknesses that could be recognized as best practices or that could lead to disparities and divides in benchmarking of open data initiatives.

Thus, two data matrices were created and loaded into STATISTICA 12.0 analytics tool. Here, standardization was first applied, and then the cluster analysis was performed. The non-hierarchical K-means clustering method and hierarchical algorithms were applied. First, the initial setup of the centers of the clusters was done using a hierarchical single linkage algorithm and Ward's minimum variance method. By checking the dendrograms for both methods, we can get information about how the clusters are formed. Thus, the non-hierarchical clustering was carried out using the K-means algorithm for matrix 1 (6, 7, and 8 clusters), and matrix 2 (5, 6, and 7 clusters). Out of the given numbers, the highest quality clustering is ensured by 7 clusters for both matrices (e.g., intra-cluster and inter-cluster distances, no empty cluster, no cluster with a single member, etc.). This number was selected for further processing. The patterns in each cluster are displayed in Table 8 and Table 9. The pattern of each cluster with the largest distance from the center is in bold.

In Table 8, in which patterns are clustered based on their similarities in occurrence across sample countries, *cluster 1* consists of patterns that suggest a link between telecommunications infrastructures and networks for services provided by open data portals and the importance of collecting and working with feedback for further development of the OGD initiative. Citizens and businesses, as well as governments, need to be informed and educated/trained about the existence of OGD, data portals, and how to reuse data. Government agencies, cities, municipalities, and other public sector organizations should also develop their own (open data) activities. *Cluster 2* can be characterized as a group of patterns that are closely related to the development of e-government, especially how OGD strategies have been developed and published over the years when official open data portals were first launched and what other websites provide data in open formats.

The patterns in *cluster 3* then suggest that if a country launched a centralized e-government citizens' portal and improved it over the years, no national reports assess/evaluate the development of open data efforts in the country or benchmark the country with other countries. These countries usually have a robust e-government system, and OGD is considered one of the services that are an integral part of the e-government strategy. Among the patterns that occur together in *cluster 5* are those that indicate that the ODE should consist of various components, especially various types of data portals that support various activities of various stakeholders that are interrelated and cooperate/interact with each other to build a resilient and sustainable ODE. This is enabled because the first open data portal was launched very early, and the users' feedback is considered in reviewing the portal when setting up the agenda for its improvement so that the ecosystem could grow. In addition, there are one or more national reports on the assessment/evaluation of a country's digital public services that support the reuse of OGD by the private sector and training for administration on opening data to improve the quality and openness of shared data and raise awareness of the benefits of making data available for reuse.

The list of patterns in *cluster 6* suggests that there is no pattern difference if the country has only one official national open data portal or there are more data portals, i.e., either unofficial or regional, local, local, city, etc. Also, portals usually occur along with other data-related activities such as feedback gathering, surveys/questionnaires, hackathons, workshops, courses, impact measurement, improving data skills, training, etc. *Clusters 4* and *7* don't provide any meaningful distinction on benchmarking open data initiatives over the years.

*Table 8.* A list of patterns in each cluster with respect to their occurrence

| **cluster 1** | A06, C01, C08, C13, C14, C15, C16, C20, C27, **C28**, C29, C30, D21 |
|---|---|
| **cluster 2** | A02, A03, A04, A07, A08, A09, A10, A12, A13, A14, A15, A16, A17, A18, A19, A20, B01, B02, B03, B04, B05, B06, B08, B11, B17, C25, D05, D06, D07, **D10**, D17, D18 |
| **cluster 3** | A01, A05, A11, C02, C03, **D20** |
| **cluster 4** | **B12**, B14, B16, C22, C26, D04, D12, D15, D22 |
| **cluster 5** | B07, C04, C07, C12, C19, **C24**, D02, D08, D13, D14 |

| cluster 6 | B09, B13, B15, C05, C09, C17, C18, C21, C23, C31, D01, D03, D09, D11, D24, **D25** |
| cluster 7 | B10, **B18**, C06, C10, C11, D16, D19, D23 |

Bold style is used to indicate a pattern with the largest distance from the center for every cluster

Ax, Bx, Cx, Dx is the pattern identifier, where {A,B,C,D} stands for the context (A - e-government, B - OGD, C - open data indices and rankings, D - other relevant resources), and x is the pattern number (see the list of patterns in Annex 2)

For Table 9, in which patterns are clustered based on their similarities in effects (impacts) on benchmarking of open data initiatives across the sample countries, *cluster 1* includes patterns with limited to average effect. Among them are all types of open (government) data strategies at the national level, i.e., (1) there is an official national OGD strategy, (2) there is no OGD strategy, but there are guidelines, best practices, recommendations, (3) the topic is only included in strategies focusing on digital technologies. Some patterns cover the lack/non-existence of national reports that assess the development of open data efforts in the country or benchmark the country with other countries. We can argue that thawing an OGD strategy is not a key element of ODE. From other patterns, it can also be concluded that e-government services, their interoperability, availability, transparency, efficiency, etc., have a positive impact here, i.e., to what extent OGD and related concepts will merge with e-government and can use its infrastructure and related services for their growth.

Patterns with no to limited effect are included in *cluster 2*. Most of them are from context *D - other relevant resources*. Their limited effect is because these patterns were found in only one or two countries in the sample, so there is no effect for other countries. However, these patterns can be a valuable source of information and best practices for other countries, and these patterns are discussed in more detail in the next section. Among the patterns from other contexts, the early government commitment to launch an open data portal and the existence of an open data portal for non-government data from business, culture, NGOs, and/or research has limited effect/impact on the benchmarking of open data initiatives. Other patterns from context D can be found in *cluster 4*, but with a limited to average effect. Among them, especially those focused on measuring the impact of open data, the importance of high-value datasets, the need for advanced features in data portals to work with datasets, and the cooperation and collaboration of all stakeholders are included. Another cluster that includes patterns with a limited effect is *cluster 5*.

In contrast to *clusters 2* and *4*, it consists of patterns from the context *C - open data indices and rankings*. The main reason for their limited effect is that open data benchmarks and reports do not assess open data portal features' existence, quality, and usability. There is no list of them that would specify them and how they contribute most to the impact that open data create etc. Also, the topic of open data showcases (use-cases/reuses/stories) and co-creation, as well as levels of stakeholder engagement, participation in hackathons, webinars or seminars, forums or online courses and other trainings, that improve skills of stakeholders are underestimated in open data benchmarks and reports.

*Cluster 3* includes patterns with an average to much effect. The most important here is a centralized, one-stop portal (one-stop-shop) that provides secure access to digital public services and a national open data portal. Both of them are affected by the legislative environment. The effects and impacts of the quality of these portals on the outputs of the respective benchmarks depend on the level of stakeholders' digital and open data skills, how informed and trained they are about the existence of

OGD, data portals, and how to reuse data. However, it should also be noted that monitoring and ensuring the use of open data is often beyond the personnel and financial capacities of the country, as well as regions, cities, and municipalities. Patterns with average effect are grouped in cluster 6. Their similarity lies in the focus on infrastructure and networks, including their support for advanced digital public services delivered to citizens and businesses, such as AI, Machine Learning (ML), Internet of Things (IoT), blockchain, etc., and the need for the management of the technical background, i.e., infrastructure, technological advances, and knowledge base.

Patterns with the highest effect (from significant/much to extreme) are included in *cluster 7*. Most of these patterns are in the context of *A - e-government, confirming the e-government system's importance for OGD*. The start of building public sector information systems and base registers, telecommunications infrastructure and networks, digital identity (eID) issuance and availability of digital public services, prioritization of security, reliability, and related policies for digital public services such as authentication, authorization, e-signatures, etc., and continuous improvement of these e-government components support the entire process of OGD disclosure and reuse and make it easier and more effective. One of the main benefits relates to the availability and interoperability of digital public services that allow OGD to be integrated into the e-government system instead of providing an isolated service.

*Table 9*. A list of patterns in each cluster with respect to their effects (impacts)

| cluster 1 | A07, A08, A12, A20, B01, B02, B03, B04, **B06**, B11, B12, B18, C02, C03, D10, D18 |
|---|---|
| cluster 2 | B07, B13, C04, C06, C19, **D01**, D02, D04, D09, D11, D14, D15, D20, D23, D24, D25 |
| cluster 3 | **A11**, C08, C09, C13, C14, C16, C17, C20, C25, C26, C27, C29, C30 |
| cluster 4 | A19, **B17**, D03, D05, D08, D12, D13, D16, D17, D19, D22 |
| cluster 5 | C01, C07, C10, C11, C12, C21, C22, **C24**, C28, C31, D21 |
| cluster 6 | A06, B08, B09, B10, B15, **B16**, C05, C18, C23, D07 |
| cluster 7 | A01, A02, A03, A04, A05, A09, A10, A13, A14, A15, A16, A17, A18, B05, **B14**, C15, D06 |

Bold style is used to indicate a pattern with the largest distance from the center for every cluster

Ax, Bx, Cx, Dx is the pattern identifier, where {A,B,C,D} stands for the context (A - e-government, B - OGD, C - open data indices and rankings, D - other relevant resources), and x is the pattern number (see the list of patterns in Annex 2)

## 5.2 Recommendations and future steps

This section reflects on findings derived from other resources (*context D*) that have been found relevant (for sample countries) and focus on developments and future steps (see Annex 1, section "Other relevant resources dealing with developments and future steps"). As such, they are not limited to any of the primary contexts we covered above but are more general, covering both those that directly affect the OGD initiative, as well as those closely related to benchmarks and reports, and thus which rather influences the results in them, as well as the overall development of the OGD initiative, including its sustainability and resilience within the country and in comparison with other countries. A total of 25 patterns were found for this context (see Annex 2). We transformed them into (high-level) recommendations expected to be taken for increased sustainability and resilience of the OGD initiatives and ecosystems that primarily rely on the best practices we came across studying selected countries, i.e., those that can be considered key to success.

**R1. Develop and maintain a multi-perspective, stakeholder analysis-centered OGD strategy**

The first set of patterns highlights the importance of having an established and well-maintained open data strategy that is seen as a set of guiding principles, a reflection of the current state of affairs. Also, a document setting an agenda for related stakeholders, as well as associated tasks and their allocation. These are used in determining the next edition of the strategy, which, in turn, is used to assess *what has been conducted so far*, *whether the expected outcomes have been achieved*, and *what corrective actions should be taken*, as well as *how the OGD initiative should develop considering the current worldwide trends and/or best practices*. While this document may be considered as having intrinsic/internal importance to those stakeholders involved in the opening and maintaining data, it is considered important to a wider audience. It facilitates establishing an understanding of the existence of the OGD initiative in a country, setting a clear vision of such, as well as serving as a reference point to initiate discussion, and is certainly considered as a good practice.

The premise/prerequisite is that the strategy must be developed and maintained responsibly, where not availability/existence of such, but availability and quality are the driving forces for the maturation of a sustainable and resilient OGD strategy. It must be compliant with a clear strategy of *what it should look like* and *how it should evolve/develop*, with a clear indication of *who is in charge of this*, and *what instrument will measure the success of completion of the above*? It also means that this strategy cannot consider a limited scope, meaning that its development and maintenance must be multi-perspective, where these perspectives are not limited to the entities and/or artifacts of the ODE, such as data and portal, but considering the wide range of stakeholders expected to be involved. This means that a significant part of the definition of an OGD strategy is a precedent and then continuously maintained stakeholder analysis. Once a set of stakeholders has been identified, further exploration of other elements of ecosystems needs to be conducted considering all stakeholders' viewpoints, needs, and expectations. None of the above can be a one-time task but a continuous set of tasks. Thus, we can think of it as a life cycle similar to the Deming; also PDCA cycle (plan-do-check-act), or define-measure-analyze-improve-control, or phases of Lean Six Sigma. This life cycle consists of at least such phases as (1) the identification of a current state of affairs and setting a list of tasks, considering the current data supply and stakeholders' needs, including an analysis of the impact and value the data create/brings, incl. their reuse and factors affecting it, data publishers involved and their capacities, etc., (2) implementing it, evaluating performance, including (3) evaluation of pillars determined before (such as stakeholders, long-term objectives, etc.), (4) take a decision on the need for adjustments to be included in the next strategy (adapted from Nikiforova et al., 2023).

**R2. Review and update the OGD portal regularly in line with users' needs and expectations and best practices, keeping the source-code and documentation accompanying it publicly available**

Another set of patterns refers to the OGD portal, which is the point of contact for data owners/data publishers and data users, and the need for regular updates and improvements of the portal. The latter is expected to be based on a list of factors including but not limited to internal audits, feedback received from (potential) users. It, in turn, implies prior stakeholder analysis, identification of these groups of potential users, their capabilities, needs, requirements, level of satisfaction with the current portal and those deemed necessary for its actual use. This also includes the study of best practices. A mechanism is needed to collect these needs. Participation in obtaining this information must be created and facilitated, which may vary for different countries.

The identified patterns set a list of prerequisites for portals, including clearly defined, well-presented, and elaborated licensing information. It is seen as crucial to the OGD, along with the need to regularly review and improve the portal, considering identified weaknesses in implemented functionality, as well as determining additional functionality that a potential user may need. This can also be affected by data supply with reference to a data type, where new data types and formats may require changes in the portal. In other words, the ability to upload and download a stand-alone file was enough with further advancement in the ability to preview a structured dataset, which is no longer sufficient. This is also related to several other patterns we discuss as part of other recommendations.

Otherwise, the OGD philosophy suggests that OGD can be used by anyone, regardless of knowledge, digital skills, data literacy, specialization, etc. This means that some user groups do not necessarily have sufficient knowledge to work with data, where it is expected that the portal will have features that support these users, thereby eliminating or at least reducing the digital divide. At the same time, providing features that support users with limited understanding and knowledge is not sufficient to facilitate its use by all users' groups. This means that more experienced users have different needs and expectations, e.g., to easily retrieve data into their application rather than preview the data and/or create a visualization in the portal. This brings us back to the need to understand potential users with the need to conduct stakeholder analysis and consider their viewpoints. It also involves the creation of various feedback channels to collect these inputs. These channels are expected to vary in nature, starting from more independent and static in nature, where the user can submit a proposal, request, or recommendation on the portal, continuing with regular surveys that can be accessed through the portal, and the continuation of more interactive workshops, hands-on and questionnaires that encourage more active participation of key and potential actors by reaching them.

In addition, continuing on the philosophy of OGD and openness in general, some patterns indicate the need to keep the list of updates of the portal and its functionality the same as the code to be open source (e.g., on platforms such as GitHub or GitLab). This is expected to contribute to the OGD movement in several ways. Firstly, by supporting the paradigm set by OGD and thereby promoting a change of mindset from a closed paradigm to an open one at all levels, i.e., not only in relation to data, by providing an opportunity to establish a feedback channel, facilitating the involvement of enthusiasts contributing to the co-creation, as well as collaborative and participatory design and development, even if through gathering active feedback and suggestions expected later to be designed, developed and integrated by the portal owner. It also demonstrates the dynamism of the portal, i.e., it is regularly maintained, and updated, ultimately leading to a more user-aligned, progressive, sustainable, and resilient portal and OGD initiative as a whole.

**R3. Develop open data literacies for different stakeholders and actors**

Open data literacy and capacities form another set of patterns where different target groups of stakeholders have been identified, namely (1) business, (2) government, and (3) enthusiasts/activists represented by citizens and academia (researchers). Relevant events and activities such as hackathons, workshops, courses, summer/winter schools, user meet-ups are expected to be organized for these stakeholders. These are expected to be regular events, held at least once a year, where groups of people with shared interests are expected to be trained and educated. This, in turn, can also be seen as one of the activities to promote OGD, as well as a feedback channel that can collect recommendations. It involves determining the form of these events that is the most appropriate for the groups and the country in question, ways to attract people to these events, as well as the content, which may also vary depending on the level of a particular cluster of representatives of the stakeholder group. As an example, for enthusiasts, these may be for those with advanced knowledge and experience, i.e., a datathon for developers and a hackathon for young people, like in Latvia, where an open data hackathon for Generation Z is organized annually, while for the government, these may be those of general nature raising awareness of the principles of OGD with public agencies, and those that cover more technical aspects. Here, it is also important to mention that while we determine three general stakeholder groups, they can be broken down into smaller groups, including but not limited to government employees, innovators/developers, data journalists, activists, and citizens.

In the case of the government as a stakeholder, another pattern is focused on the need to train public agencies and their employees on opening data. This aims to improve the quality and openness of data and raise awareness on the benefits of making data available for reuse, thereby increasing interest in data, and in particular high-value data, opening, and maintenance as a source of innovation and social and environmental value. Moreover, being more related to another recommendation we will present below, it is indirectly related to the possibilities of providing feedback to users regarding requested data. In other words, for those requests for datasets to be opened, where the user is expected to receive feedback on whether the dataset in question can be opened, and, if not, what is the reason, i.e., non-compliance with OGD principles or similar. The latter is, unfortunately, a rare practice, although it has proven to be very useful as a source of education to increase open data literacy level, and to build trust and facilitate users participation in the OGD initiative. This is particularly important, considering that open data literacy is seen as a game-changer to the problem of open data usage.

**R4. Establish a national hub/center to support the public administration in using technological advances**

Considering a strong relation to and dependence of the OGD's success on the ICT advances, public administration is expected to be aware of, sufficiently experienced, and familiar with them. As is the case for all types of data and is not an exception for open data, AI is one of the most expressive examples of this technology, which was observed as one of the most important patterns. However, considering its complexity and limited capacities and resources within public administration, this pattern suggests that an establishment of a national AI hub/center that would support the public administration in using AI in an ethical, robust, reliable, scalable, and secure manner could be the key for success.

This is even more so since it has been observed that countries that prioritize AI, promoting its wide adoption, particularly in public administration, but also at other levels, see great value in OGD. This is also compliant with the current discussions around these two topics and especially their combination, where both concepts set the prerequisites for the development/maturation of both of them, i.e., *AI for OGD* and *OGD for AI*. Similarly, but not as often as in the case of AI, the call for developing a digital or information society is seen as the key to success; thus, establishing centers for developing digital and data literacies is expected to bring benefits.

Additionally, indirectly related to it emphasizes the importance of establishing interoperability and integration. Their support is expected to reduce the administrative burden associated with providing services to citizens and businesses, as well as the importance of providing guidance and manuals for data providers (ministries, regions, cities, and other stakeholders), i.e., open source software, pattern labs, search engine optimization, etc., intended to help other teams create digital products faster (websites and applications) that will be consistent across the public sector agencies and institutions. The above, however, are related to digital government and digital society, which as a result, will have a positive impact on the OGD initiative.

**R5. Establish an interaction and long-term cooperation with the community and ecosystem stakeholders**

One of perhaps the most important, though obvious, patterns is the need to establish interaction and cooperation with the community and ecosystem stakeholders. This is expected to be a key prerequisite for a sustainable and resilient OGD ecosystem and OGD initiative as a whole since, to exist and remain sustainable, stakeholders must be involved in the OGD initiative at all levels. However, identifying these stakeholders and how they can be reached and involved in the OGD initiative is challenging. This is where stakeholder analysis can be used in combination with R3. Again, once this communication has been established, it needs to be maintained regularly, including determining new stakeholders. This, in turn, will contribute significantly towards communicating the current state of affairs and its relevance to real needs, collecting feedback, and identifying a list of improvements, aligning it as much as possible with the needs and expectations of stakeholders. This would contribute to the creation of value within the OGD initiative, and serve as an input for every next edition of the OGD strategy that will make sense instead of being a stand-alone document that does not bring any or a very limited added value.

**R6. Define, open and maintain high-value datasets**

While the OGD strategy, the OGD portal, its features, and citizen engagement make sense, they will not matter if open data are not of interest for reuse. In other words, data availability should not in itself be a goal, where the availability of data of interest to citizens, businesses, and any other stakeholder is important. This is especially important for data that may be of interest either to broader stakeholder groups or those that have great economic, social or environmental potential, namely the notion of High-Value-Datasets (HVD) introduced by the Open Data directive (previously Public Sector Information Directive)[3,4].

This, however, is an ongoing topic, where although Open Data Directive has taken steps to define both HVD categories and a list of specific datasets, they (1) are general in nature and given the content can be considered to be opened by default, (2) focus on an increased harmonization and the cross-border interoperability of public sector data, and data sharing across EU countries. The latter, as a result, leads to the case that they are of high international or European importance, and national importance, however, does not consider those datasets that are relevant for a particular region and/or country, its society, economy, and environment. Thus, countries are expected to determine country- and region- specific HVD themselves. It is not clear how this can be done since there is no universal

---

[3] Directive (EU) 2019/1024 of the European Parliament and of the Council of 20 June 2019 on open data and the re-use of public sector information (recast), https://eur-lex.europa.eu/legal-content/EN/TXT/?uri=celex%3A32019L1024
[4] Commission Implementing Regulation (EU) 2023/138 of 21 December 2022 laying down a list of specific high-value datasets and the arrangements for their publication and re-use

framework and/or approach for this yet, but rather ad-hoc approaches are currently being adopted in some countries (Nikiforova et al., 2023).

In addition, as regards the definition of generally valuable or country-specific HVD, it is expected that the national portal should allow users to request the dataset and track the status of the request transparently (and receive feedback on a reason why they cannot be opened, if it is the case). There is an expressed need for support for releasing valuable datasets (of national value, i.e., country-specific) as open data. Here, several categories are expected to add value to the OGD initiative by making it more relevant and aligned with the current needs. They are geospatial data and/or real-time or dynamic data, which are also consistent with the general definition of the HVD term, as well as a new category of data considered to have a high potential for sustainability and resilience of OGD ecosystems - citizen-generated data, whose publication and access to which is expected to be incentivized by national strategy/policy.

Of course, these recommendations can be seen as very interrelated, and when seen together, more specific detail may emerge from these high-level recommendations. In other words, some of the aspects that are mentioned as part of the more behavioral recommendations can also be translated into technological solutions, e.g., features and tools for the portal and vice versa. To sum up, they can be characterized as recommendations oriented towards *data*, *portal*, *people* – providers and consumers, *capacities and skills*, *citizen engagement* and *communication*, *feedback channel*, *impact*, and *data value*. All should first be understood, established, and then maintained, creating a complex and dynamic sustainability-oriented ODE.

## 6 Discussion and limitations

### 6.1 Discussion and recommendations

High expectations from OGD are followed by a series of setbacks and a realistic recognition of the hype, challenges, and limitations, especially in achieving development outcomes and impact (Meuleman et al., 2022). The findings of our study are of great importance for individual countries, i.e., of national importance for eight countries. These findings allow a more accurate and correct interpretation of results and changes over the years within a particular index or rank (at least six of those we covered), i.e., whether the difference in results is the result of national efforts or the subject of changes in a particular index. Also, how to combine and interpret the results of several indices for correct decision-making and for defining future actions when the results differ significantly (Lnenicka et al., 2022).

While changes in benchmarks and indices methodologies, same as the metrics used and their weights, change over time and may lead to misinterpretation of the development of the OGD initiative, considered a risk when country development is evaluated based on them, we have found that this is not necessarily true. In other words, very few countries rely on existing benchmarks and indices, and in particular track progress over time and set their agendas based on them. This does not mean that these benchmarks and indices are not used, as they are used, referred to in national OGD strategies or other documents used instead. Still, they mostly refer to the active edition when the strategy was defined, with less attention paid to the progress of the OGD initiative in these resources over time. Some countries that have been found to ignore existing benchmarks and indices reflecting on OGD initiatives in multiple countries (with European or even more global scope) use alternative ranking and reports, e.g., national or regional level. This, in turn, was found to be due to the geographical, economic, cultural proximity that tends to affect the e-government and OGD behavior. However, it was found that they are often based on or follow methodologies established by those widely adopted ranking and indices, with a more detailed and in-depth analysis of a country in question.

This, however, may be due to the fact that existing benchmarks and indices tend to be of a high level or provide a limited overview of either the current state of affairs, information on what and how can be improved, or why a particular aspect is assessed as low quality, which limits the ability for root-cause analysis the countries expect (ODMR, however, tries to fix it). At the same time, some benchmarks, such as ODMR, rely on data that are first provided by the OGD initiative holder - self-assessment, which can be intentionally or unintentionally falsified due to the pressure to look better compared to others that increase the risk for the data collected in this way to be unreliable (Bannister, 2007; Lnenicka et al., 2022; Nikiforova and McBride, 2021). Although the data reported are then checked, it is clear that in-depth analysis is too resource-consuming, where resources are not only about money or time but also about human resources. More importantly, in addition to the people expected to be involved, the context and in-depth understanding and knowledge of the OGD initiative, the OGD ecosystem, its elements, and relationships are necessary for the country in question, who in addition to the above should be independent, which does not seem to be feasible.

Additionally, most of the assessments are binary (0 - not fulfilled, 1 - fulfilled), despite the fact they are not. I.e., if the person filling out the protocol indicates that something is done. This, as a rule, is not only not checked, but also the highest number of points is assigned both for a perfect implementation, and for an unsuccessful or a poor attempt, or just an attempt to get points. Thus, benchmarks and indices relying on this data collection tend to provide insufficiently accurate results. During the study, we also noticed that when data incompatible with the actual state of the art are provided, this is not necessarily the case when a non-existent positive result is reported - in some cases, positive results existing in the country are not reported (possibly due to lack of awareness, knowledge of a person in charge of providing these data).

**R7-14:** As a result, several recommendations can be defined arising from the above. First, **benchmarking developers/producers** are encouraged to:

> (1) check the reliability of the data provided to them, particularly if it is done by the OGD initiative owner (**R7**);
>
> (2) make sure that more than one ODE stakeholder group is involved, in particular, to make sure that not only data provider perspective or OGD initiative owner is the only source of data (**R8**);
>
> (3) consider the quality and maturity of the fulfillment of certain criteria instead of using a Boolean assessment that ranks those that provide their users with a qualitative and user-friendly service or product the same as for those, who rather strive towards higher scores and at times a very limited focus (if any) on the needs, expectations and level of satisfaction of the end-user merely formally fulfilling the requirements (**R9**).

At the same time, several recommendations **for benchmarking users**, incl. ODE owners are:

> (1) when interpreting results obtained in a benchmark, report, or index, examine the methodology and indicators used (**R10**),
>
> (2) when interpreting and comparing country results compared to previous years, study the changes in methodology first, which are typically the subject for changes for every benchmarking. This as a result affects the rank and the points obtained, which is not necessarily an indicator of progress or failure (**R11**),

(3) do not take the result, incl. in comparison with other countries, for granted, where the weights of each indicator obtained and the method of scoring should be examined, making sure whether only the formal fulfillment of the requirement is assessed, or the quality of its implementation is also assessed (**R12**).

And finally, a few more recommendations **for ODE owners**:

(1) when submitting data required by an index, benchmarking or report developer, ensure that the data are accurate, reliable and compliant with the actual state of affairs checking them. They are also expected to consult respective bodies / stakeholders and actors, who can provide more information, e.g., academia (**R13**);

(2) do not rely on external (international) benchmarking, indices and reports. Given the complexity of gathering a large amount and diversity/variety of data, along with the complexity of their verification and general purpose, they tend to provide a high level understanding of the state of affairs in the given area, with a limited understanding of the internal situation of the specific OGD initiative in question. Hence, examinations of the ODE and OGD initiative at lower level should complement the above sources (**R14**).

Alternatively, other indices and benchmarks, designed to be more independent and objective, which do not suppose the involvement of representatives of the OGD initiative in question, although they somehow resolve the above issue, still tend to lead to false results. This is due to a lack of deep understanding of the OGD initiative in the country, i.e., remaining unaware that the data source has changed and the one used previously is no longer maintained, resulting in low results in these indices, but does not reflect the actual state of affairs in the country (the case for ODIN). This leads not only to inaccuracies in these results, but also to the failure to later use these findings for the purposes for which indices and benchmarks are intended.

Both, i.e., either the knowledge of reporting incompliant data, or observing results that do not correspond to the actual state of affairs, can also cause resistance/reluctance/unwillingness to use these indices and benchmarks. This may be the reason for the low use of these indices observed in the countries we analyzed.

**R15-16:** Hence the recommendation **for the user**: if the index, report, or benchmark collects data independently of the open data initiative owner or any other representative of the country in question, when interpreting the results, make sure that the source used to collect the data is one that is used in the country - seeking for increased objectivity of the results indices may isolate themselves from factors that may influence the result, but may end up relying on sources whose analysis is irrelevant as obsolete or replaced by other sources (**R15**). The recommendation **for ODE owners**, in turn, is to check the same and timely identify issues implying incorrect and inadequately low results due to the use of an outdated or simply wrong data source, which, as a consequence, negatively affects the overall image of the OGD initiative and the country (**R16**).

While there are many open data-related indices and benchmarks, ODMR is the most widely used, probably due to the fact it is a very detailed (the highest number of variables and the one of the only indices that cover impact), multi-perspective, and the best maintained overview. Concerning the latter, maintenance, this can be seen as one of the decisive factors why countries prefer it, since, when other benchmarks and indices were regularly updated, ODMR was not only widely used. Another reason for its popularity among countries, especially within the EU or candidate countries, is that the ODMR can be seen as a guidance for making an OGD initiative compliant with general EU and European Commission requirements regarding the OGD initiative. This is due to the fact that the ODMR is also regularly revisited to align it with current and future OGD trends. This is intended to make the methodology oriented on increasing sustainability and resilience of OGD initiatives, as well as to ensure that national OGD strategies are kept up-to-date and ready for upcoming changes introduced by the OD Directive. It is also an educational source that provides an overview of best practices in many subdomains from which OGD initiative holders can learn. However, while many countries refer to the ODMR in their national OGD strategies or other relevant documents such as Smart City or Sustainable Development strategies, most mention them rather than design the agenda and next edition based on it. Moreover, in many cases, it is referred to if these results are complaint with the overall strategy and vision of the OGD initiative holder, similar to what Selten et al. (2023) found for the trust in AI recommendations by "street-level bureaucrats" that occur if these recommendations confirm their judgment, what they call "*Just like I thought*." In other words, when these findings can support and perhaps manipulate further actions, even if they are not consistent with other findings, e.g., public perception or the results of other benchmarks or indices.

Many countries, however, seem to have a greater preference for more e-government-oriented benchmarks. Benchmarking in the public sector, which can sometimes be considered e-government benchmarking, mainly concerns policy makers, who can be seen as their target audience (Heeks, 2008). In this direction, benchmarking is a retrospective achievement (tracking the results of a country or agency in certain rankings), as well as promising direction/priorities (to achieve high performance in e-government). The most widely recognized e-government and ICT-related benchmarks in the public sector are the UN-DESA with the EGDI (first published in 2001, the latest report is from 2022), the World Economic Forum with the Network Readiness Index (first launched in 2002, the 2019, 2020, and 2022 editions are grounded on the Portulans Institute), and the International Telecommunication Union with the ICT Development Index, which was published annually between 2009 and 2017 (Bannister, 2007; Machova and Lnenicka, 2015). Ki (2021) stresses the importance of learning of local government officials from their peer governments. In this way, from successful managerial experience, valuable conclusions can be drawn, which can further help government officials for more quality activities to serve the public. The causal relationship between benchmarking as a reflexive institution and the actual innovation capacities of governments enables the identification and explanation of institutional differences. They impact the benchmarking process as a reflexive institution through three features of benchmarking, namely - obligation, sanctions and benchmarking authorities (Kuhlmann and Bogumil, 2018).

**R17-19:** Recommendation for the **OGD initiative owner**: use well-maintained indices, benchmarks, and reports to determine *what the current state of affairs is*, *how well or poorly it is performing compared to other countries*, identifying from the above corrective actions, and use them as a preliminary input for an in-depth analysis and OGD strategy or relevant document (reducing the resources to be spent on their formulation), *what are the best practices*, as well as *what are the current trends to which the OGD initiative should be adapted*, or *what preparatory tasks should be launched*, considering current advances associated with the OGD (**R17**). Combine multiple indices, benchmarks, and reports to derive as many insights as possible (**R18**). Considering the best practices reported in the above documents, think about your practices and examples, approaches to identifying them (in case they can exist), reporting them (if they exist), or promoting their implementation in the country (**R19**).

It also happens that some countries derive recommendations or identify pain points from the results obtained in the OGD benchmarks and indices. However, what is interesting, is not necessarily related to the country's ranking in a particular benchmark. We found that countries can be divided as:

(1) rank high and work hard to develop in line with their recommendations;

(2) rank high but do not care too much about what these OGD indices recommend having their mechanisms in place;

(3) rank low and do not care about results and recommendations;

(4) rank low but work hard to improve results based on the OGD indices;

(5) rank low but do not care too much about their recommendations having their mechanisms in place.

The fact here is that countries that rank high are often making more attempts to maintain this position, which is most likely not so much about ranking but rather about maintaining the OGD ecosystem and the entire initiative. This behavior, however, tends to change, as some countries ranked high at some point, but then began to lose their positions and have never returned to high positions until now. In the case of the sample countries, this was suggested due to the fact that at the beginning of the rapid development of the OGD initiative, there was substantial EU funding, including for the development and launch of the portal, but when several thresholds were reached such as those related to the launch of the national portal and some progress in data provision, maintenance activities did not occur, but occur by inertia and/or based on the efforts of enthusiasts (also with reference to data publishers), when they can contribute.

**R20:** Recommendation for the **OGD initiative owner**: the OGD initiative is not a once-only process where it is expected to be established. Instead, it is an ongoing continuous process, where the OGD initiative and the entire ODE ecosystem should be continuously maintained, regardless of the current result, i.e., it should be improved continuously in both cases if it performs poorly or is considered the most competitive (**R20**).

Universal implementation tools for open data, which are frequently spread regardless of various institutional contexts, should be updated and reclassified in accordance with long-standing customs in public administration at various institutional levels, especially in countries with various related decision-making and governance mechanisms. These solutions and standardized digital platforms should be categorized in accordance with established ODEs. In this regard, the related methodologies should be updated such as global indices as the biennial global e-government survey from the UN, the GODI from the OKF, the OGI from the World Justice Project and many other assessing methodologies (Kassen, 2018).

**R21:** An understanding of what constitutes the ODE is required, with reference to its components, actors and stakeholders, their roles, and relationships between the elements of this ecosystem. This is expected to be addressed primarily by academia in collaboration with other ODE stakeholders, particularly public administration. When defined, it should be maintained with a focus on keeping it up-to-date, resilient, and sustainable (**R21**).

It is also important to understand that open data is based on the concept of open government, that is, on a subset of the concept of e-government/digital government. This can be seen as an obvious fact since open data is one of the integral elements of the open government initiative. Nevertheless, Lnenicka et al. (2022) argue that benchmarks of open data efforts should be viewed through other elements, not just through open data. Elements such as e-readiness, ICT capacity, ICT preparedness/readiness and ICT penetration form the ODE, where the overall state of each determines the value of indices and rankings of open government efforts in various aspects, thereby showing the commitment of government authorities to the principles of open government.

One more important issue results from the ecosystem approach and how relevant stakeholders are supported to communicate and interact with each other. Without a clearly defined structure of components and responsibilities of its actors, the ODE cannot provide an environment in which value can be created from open data. This limits the actions of governments because the supply and demand sides are poorly coordinated. As stated by Welle Donker and van Loenen (2017), open data assessment frameworks do not have to cover all parts of ODEs. Thus, it is important to know how these ecosystems are structured, their specifics, components, and relationships, and then apply the most suitable tool to get relevant results.

## 6.2 Limitations

This study has some limitations, one of which results from the use of the cross-country study approach. While cross-country comparative research was the only viable option for gaining an in-depth understanding of the most relevant patterns of different countries in a field *"based mostly on observations, expert opinions and experiences, previous practices and aggregate data"*(Gharawi et al., 2009, p. 5), such a methodology comes with risks associated with data analysis and the research process.

Data analysis in cross-country case studies is subject to issues of data collection and reliability. Data collection in cross-national case studies that rely on experts, as in the Delphi methods used in this study, is based on input provided through the initial questions (Franklin and Hart, 2007). Therefore, it is important that the questions fully and truly reflect the research objective of the study. To this end, our protocol attempted to capture all the development patterns relevant for the research objective of this study through a multi-stepped process in multiple rounds. Although every effort has been made to ensure that the input contexts reflect the research objective of the study, there is still the possibility that some elements may have not been included in the data collection.

In terms of data reliability, due to the multicultural and multilingual fashion of such studies, data can have different meanings and interpretations in different countries (Gharawi et al., 2009). In connection with the previous point, another risk is related to the role of the selection of experts in the Delphi method. Indeed, the reliability of data depends on the perspective and knowledge of the experts and, as such, their purposeful selection is vital to the results of the study (Franklin and Hart, 2007). In our study, we dealt with these risks by developing and testing a protocol with refined and clear wording, and by selecting a panel of knowledgeable experts compliant with the predefined expert profile. At the same time, the risk associated with the multicultural and multilingual nature of the study was reduced by the involvement of experts representing selected countries, where each expert was responsible for collecting data about the country with which (s)he is associated with.

Finally, the value of the research process in cross-national studies depends on a clear definition of the research objective, the unit of analysis, and the selection of countries for comparison (Gharawi et al., 2009). For this reason, our study relied on a rigorous process that moved from the definition of the research objective to the selection of countries for comparison. Nonetheless, we acknowledge the intrinsic sensibility of the study results to these choices.

Another limitation to consider is the existence of other indices and rankings that are not directly focused on open data and open government but include sub-indices or sections that deal with this topic. For instance, the Digital Economy and Society Index by the EU (which was found to be mentioned in OGD strategies of sample countries) includes open data as part of the digital public services indicators. But these data come from the ODMR, which is an OGD index on its own. The question is whether countries prefer these more complex/sophisticated reports and indices as they provide a comprehensive view of the overall area of digital transformation, e-readiness, or e-government development.

# 7 Conclusions

This paper aimed to explore and understand how various open data benchmarks evaluate and compare countries over the years. It is obvious that each framework used for this purpose has its goals, sets of indicators, methodology, etc., which affect the position of the country in the rank. This decreasing or increasing position is more or less influencing how governments respond to them.

Thus, we have attempted to identify patterns observed in key sources, with further evaluation of their impact, which may lead to disparities and divides in the development and benchmarking of ODEs. Thus, we identified existing benchmarks, indices, and rankings of open (government) data initiatives, further analyzing their scopes and characteristics. As a result, we (1) supported our assumption that there are at least several contexts that determine the success of an OGD initiative, as well as shape its development and affect its position in the rankings and benchmarks, namely, the OGD itself, including the OGD strategy and the OGD portal as the central point of communication of the OGD ecosystem, open (government) data indices and rankings, and e-government, as well as (2) identified six major open data-related benchmarks and indices, namely, GODI, ODB, OURdata Index, ODIN, ODMR, OGDI, which formed the sample we used to answer the *RQ1 - What are the patterns observed in open (government) data initiatives over the years?*

As part of the RQ1, we first selected eight sample countries to be investigated in detail, with further analysis of their specifics and performance over the years in the indices and benchmarks we identified earlier, covering 57 editions of OGD-oriented reports and indices and e-government-related reports (2013-2022) - UN E-Government Survey, eGovernment Benchmark, which were then supplied with UN's Economic and Social Council's Working Group on Open Data, The European Commission's policies on open data, Meetings of the OECD Expert Group on OGD, OECD Policies & Working Papers and other relevant sources of more regional and national level. They shaped a protocol (Annex 1) completed for each of the sample countries, based on which we then identified 102 patterns obtained as a result of an expert panel assessment conducted by eight involved experts in the Delphi study.

This served as input to the final RQ2 on the impact of identified patterns that may lead to disparities and divides in the development and benchmarking of ODEs, where a final number of 94 patterns was obtained representing four contexts - e-government, OGD, open data indices and rankings, other relevant resources. We then performed the cluster analysis to find similarities between patterns based on their occurrence and effects (impacts). Both these analyses suggest a close link between approaches to benchmarking of open data initiatives and the development of e-government over the years. We found that e-government services, their interoperability, availability, transparency, efficiency, etc., have a positive influence here, i.e., to what extent OGD and related concepts will merge with e-government and can use its infrastructure and related services for their growth. Finally, we were also able to extract from the 25 patterns six high-level recommendations that are considered the key to success, i.e., for a sustainable and resilient OGD initiative. The discussion, in turn, allowed us to formulate 15 more recommendations for public administration, those who use/interpret indices, benchmarks, and reports, and academia, indicating some research agenda.

These are expected to lead to improved performance in applied indices and rankings and, more importantly, will facilitate the achievement of the benefits with which open (government) data are associated. While this is expected to be primarily important in instructing ODEs' stakeholders (mainly policymakers), the findings identified the current research gaps to be further explored by researchers. As future research, we will expand the study to other countries, focusing our attention in specific areas of the OGD ecosystems to get valuable insights concerning OGD strategies used and in identifying development stages in OGD.

**Funding/Acknowledgement:** This research has been (partly) funded by the institutional support of the University of Pardubice and by European Social Fund via IT Academy programme, University of Tartu (Estonia). This research was funded by the Regional Government of Andalusia, Spain (Research projects number P20_00314 and B-SEJ-556-UGR20).

# ANNEX 1. Protocol

Country (name):

| General data | | |
|---|---|---|
| Data type | Value | Reference |
| Population | | |
| Population density | | |
| Area | | |
| GDP (nominal) | | |
| GDP per capita | | |
| EU Member / since when | | |
| E-government context | | |
| Currently valid reports<br>*Instructions: How developed is the e-government system in the country? What are its strong features, and what are its shortcomings? Please, reference to UN E-Government Survey 2022 and eGovernment Benchmark 2022, or any relevant national reports. Focus on current rank, projects carried out in the country, and other achievements.* | | |
| Findings: | | |
| Past reports | | |

*Instructions: Reflect on the progress / development over the years, emphasizing any "milestones" (if any). Please focus on the past reports of the UN E-Government Survey (10 reports between 2003 and 2020) and eGovernment Benchmark (19 reports between 2001 and 2021) series, as well as relevant national reports.*

Findings:

## Open (government) data context

OGD strategy – development over the years
*Instructions: If there is any, when was the first OGD strategy published? If not, provide a comment on why not. Has it been later improved / updated, or are there any follow-ups strategies? Is it linked to other e-government / digital strategy documents? Focus / reflect on the main principles / goals / actions / responsibilities / measurements included in the document(s). Also, include the context of resources (financial, funding, people, data infrastructure etc.) for opening data and how the availability of these resources affects the development (strategy implementation). Finally, explore if the strategy includes and supports stimulation of OGD use by stakeholders (fix how it is supported), and which of them are preferred / participate the most.*

Findings:

OGD portals – development over the years
Instructions: Consider ONLY national / country-level portals.
If there is any, when was the first OGD portal launched?
Were there any efforts (portals) before the official (national) portal was launched?
If there are more OGD portals, describe them all – why are there more portals, how do they overlap or relate, what are their differences etc.
How the portal(s) has evolved over the years – please reflect on:
  (1) number of datasets and categories (if this information can be obtained), e.g. at the time the portal was launched and current data. Can these data/ information be easily obtained? (is this information available in the form of articles / news posts, reports, open dataset, workshop where it is presented and demonstrated etc.). In case you can collect these data at a higher granularity / level of detail (different from the current data and at the time the portal was launched), please do so.
  (2) new features (whether the features have been improved? whether the portal has remained the same? were additional features added and the portal is "active"? were some of them were disabled – if yes, is the reason known, i.e. no one used them?). Can these data / information be easily obtained? (is this information available in the form of articles / news posts, reports, open dataset, workshop where it is presented and demonstrated etc.).
  (3) new data providers and other stakeholders joined the portal, e.g. at the time the portal was launched and current data. Can these data / information be easily obtained? (is this information available in the form of articles / news posts, reports, open dataset, workshop where it is presented and demonstrated etc.). In case you can collect these data at a higher granularity / level of detail (different from the current data and at the time the portal was launched), please do so.
  (4) evaluations performed to improve the quality of the portal (usability, accessibility etc.), the quality of datasets (metadata), and the metrics tracked.

Findings:

## Open data indices and rankings context

*Instructions: Refer to the link with values and evaluate the progress of your country based on the shared file with all indices. Then, refer to the reports available in the shared folder, search for the mentions about your country and analyze the development (subjective overview). Focus on the structure and components of each index / rank to understand strengths and weaknesses of the country through the years. As an expert, compare these results with your knowledge about the progress of open data efforts in the country.
How to structure your findings: analyze each index / rank separately, analyze the progress over years, compare the findings with other index / rank etc.*

Findings:

Find other assessment and evaluation frameworks – if any exist – other different resources used by the national government to comparatively evaluate the development of OGD, e.g. national annual reports or international reports that may or may not be based on existing indices and rankings, but benchmark the country (to other countries) and provide some insights into your country's progress.

| | |
|---|---|
| Findings: | |
| Summary and recommendations: Focus on the disparities, i.e., in which the country is better or worse than others, and what are the reasons / causes (in your opinion). | |
| Findings: | |
| **Other relevant resources dealing with developments and future steps** | |
| *Instructions: Which activities arising from the benchmarkings and your country's position in them are expected as future steps (strategies) of your country's development in this area? (if known) You can extract this information from the reports analyzed above, or search for other resources that are relevant to benchmarking open data strategies, development, future projects and national / global trends etc. (may not be directly related to the country), e.g., https://unstats.un.org/open-data/ (https://unstats.un.org/unsd/statcom/51st-session/documents/2020-26-OpenData-E.pdf), https://www.oecd.org/gov/digital-government/open-government-data.htm (https://www.oecd.org/gov/digital-government/7th-oecd-expert-group-meeting-on-open-government-data-summary.pdf) or https://digital-strategy.ec.europa.eu/en/policies/open-data* | |
| Findings: | |

# ANNEX 2. A list of patterns

| A01 | Continuous improvement of a centralized e-government citizens' portal. |
|---|---|
| A02 | Increasing the availability of mobile apps provided by the public sector and the use of these apps by citizens and businesses to communicate and exchange information and data with public sector agencies and institutions. |
| A03 | Development of a national cybersecurity strategy for the public sector to ensure a high level of network and information security and solutions to protect/defend against today's threats. |
| A04 | Ensuring security of operations in the public sector, new and improved tools for authorization and authentication of citizens. |
| A05 | Existence of a centralized e-government citizens' portal. |
| A06 | Improvements in telecommunications infrastructures and networks to support advanced digital public services delivered to citizens and businesses, such as AI, ML, IoT, blockchain etc. |
| A07 | Increasing engagement of citizens and their participation. |
| A08 | Increasing interoperability and availability of cross-border services. |
| A09 | Increasing interoperability of services. |
| A10 | Increasing number of public digital health services and their users to ensure resilience and competitiveness in the future. |
| A11 | Launch of a centralized one-stop portal (one-stop-shop) providing secure access to digital public services. |
| A12 | Transparency on data collection, incl. the usage of cookies and informing the user about the data expected to be collected and requesting the users' consent for their collection. |
| A13 | Launch of the public administration portal and portals of public sector agencies and institutions with relevant and up-to-date information and life events that help citizens and businesses get the necessary information online. |
| A14 | The availability of mobile apps and access to digital public services from mobile phones (in general), including the usability and friendliness of these apps, resulted from the penetration of mobile phones among citizens and businesses. |
| A15 | The start of building public sector information systems and base registers to enable efficient flow of information and data between public sector agencies and institutions. |
| A16 | The start of building telecommunications infrastructure and networks enabling access to the Internet as well as digital public services for all stakeholders. |
| A17 | The start of digital identity (eID) issuance and availability of digital public services that can be used in this way. |
| A18 | The start of prioritization of security, reliability, and related policies for digital public services such as authentication, authorization, e-signatures etc. |
| A19 | The start of prioritizing digital contact of citizens and businesses with government services as a primary channel. |
| A20 | The start of using cloud computing services to reduce / decrease costs and improve the efficiency and effectiveness of public sector agencies and institutions (in terms of hardware and software). |
| B01 | Even if there is no OGD strategy, there are guidelines, best practices, recommendations etc. on the topic of OGD provided by national or international organisations – adapted and updated for the country. |
| B02 | Even if there is no OGD strategy, there are guidelines, best practices, recommendations etc. on the topic of OGD provided by NGOs, academia – adapted and updated for the country. |
| B03 | The legislative environment of OGD is affected by the Open Government Partnership membership and related action plans. |
| B04 | The OGD as a topic is included in the national digital strategy and/or strategic documents, Action Plan dealing with digital technologies and their use – usually updated every few years. |
| B05 | The PSI Directives and the Open Data Directive by the EU are implemented into national law, usually in the context of free access to information rights. |
| B06 | There is an official national OGD strategy that is updated on a regular basis every few years. |
| B07 | An open data portal for non-government data from business, culture, NGOs, and/or research is part of the open data ecosystem. |
| B08 | There co-exist many local catalogues at the regional and local levels, open data portals of ministries, and other public sector organizations. |
| B09 | With the launch of the national open data portal, some of the unofficial, regional, or local portals disappeared. |
| B10 | With the launch of the national open data portal, some of the unofficial, regional, or local portals have been merged into the national portal but are still online. |
| B11 | Before the launch of the official national open data portal, there were decentralized public sector agencies and institutions' websites / portals with sections / tabs devoted to the publication of open data. |
| B12 | Currently, there is one portal in the country that is centralized and includes datasets of national, as well as regional and local levels. This is the only portal and the country doesn't have a local or regional portal. |
| B13 | Currently, there is only one portal in the country, which is centralized and includes datasets of national, as well as regional and local levels. It also has local or regional portals, where these data are also available. |
| B14 | The national open data portal is less than 10 years old (but 5 or more years). |
| B15 | The national open data portal is less than 5 years old. |
| B16 | The national open data portal was developed / launched as a part of an EU funded project. |
| B17 | The National Statistical Office offers a range of official statistics (datasets), following the Open Data principles, on its website / portal, and is usually an important element of the open data ecosystem. |

| ID | Description |
|---|---|
| B18 | The national open data portal is more than 10 years old. |
| C01 | There are co-creation and collaborative approaches / channels to capture the needs, expectations and recommendations of users that influence the further development of the OGD initiative. |
| C02 | There are no national reports that assess / evaluate the development of open data efforts in the country or benchmark the country with other countries. Instead, these countries rely on and plan their open data strategies based on existing open data reports, usually ODMRs. |
| C03 | There are no national reports that assess / evaluate the development of open data efforts in the country or benchmark the country with other countries. No open data reports are considered. |
| C04 | There are one or more national reports on the assessment / evaluation of a country's digital public services, digital projects, or e-government services and the topic of open data evaluation is included in the report as one of its sections. |
| C05 | There is a series of national reports evaluating / reflecting on a country's open data efforts, usually published annually by government bodies or NGOs for the government to help plan open data projects and strategies. |
| C06 | There is a series of reports on the evaluation of digital public services, digital projects, or e-government services of several countries or a region (e.g., German-speaking countries, Nordic countries), which is usually published annually, and the topic of open data is included in the report. |
| C07 | Active government engagement with the open data community. |
| C08 | Not all datasets can be previewed, e.g. in a tabular form. |
| C09 | Showcases / use-cases / re-uses / stories are not provided on the portal. |
| C10 | Showcases/ use-cases / re-uses / stories are provided and can be uploaded by the data user. |
| C11 | Showcases/ use-cases / re-uses / stories are provided on the portal WITH the link between these showcases and the data, i.e. it is not possible to identify which datasets and how were used within this showcase ⇒ the value of the data can be determined and inspire others. |
| C12 | Showcases/ use-cases / re-uses / stories are provided on the portal, however there is no link between these showcases and the data, i.e. it is not possible to identify which datasets and how were used within this showcase ⇒ the value of the data cannot be determined and inspire others. |
| C13 | Stakeholder feedback is not collected (not about data, data quality, portal functionality, expectations, reuse etc.) and suggestions from enthusiasts are often ignored (probably due to limited resources needed to make changes). |
| C14 | Stakeholders – business and citizens are often either unaware of the existence of an OGD (portal), or unaware of or critical of the benefits of an OGD closed ecosystem. |
| C15 | The datasets are accompanied with the metadata, described, and updated regularly, but the level of openness of the datasets is low (i.e., 1 to 3 stars according to a 5-star scheme). |
| C16 | The government agencies are unaware of showcases/ use-cases / re-uses / stories of their data. |
| C17 | The government agencies are unaware of showcases/ use-cases / re-uses of their data and do not make attempts to gather these data. |
| C18 | All engagement and participation is limited to ministries, public agencies and institutions – all feedback, questionnaires, hackathons etc. are focused only on these stakeholders. |
| C19 | The government's early commitment to launch an open data portal. |
| C20 | The national open data portal provides support and guidelines for data reuse, but monitoring and ensuring the use of open data is often beyond the personnel and financial capacities of the country as well as regions, cities, and municipalities. |

| ID | Description |
|---|---|
| C21 | The national portal, which previously served as a simple catalogue of metadata records, has expanded its role to include interactive information (such as showcases / re-uses / use-cases, data visualization or transformation, storytelling, feedback loop, incl. forums) and education (such as training materials, webinars or seminars, hackathons). |
| C22 | The open data strategy that considers providing linked data and ensuring interoperability between all datasets to improve their usability and facilitate reuse. |
| C23 | The technical background, i.e., infrastructure, technological advances, and knowledge base, is being developed by the National Open Data Coordinator in collaboration with academics (universities). |
| C24 | The topic of open data has largely failed to become institutionalized across government agencies within the country (especially at the regional and local level) - a very decentralized organization of the government, where agencies rely on a high level of autonomy. |
| C25 | Digital skills and open data skills in particular are lacking for citizens. |
| C26 | Digital skills are lacking for all groups, and universities rarely are positive about teaching these skills, although individuals and even groups tend to emerge, while many of them seek their own profit out of this. |
| C27 | Digital skills are lacking for public officials. |
| C28 | Geospatial datasets cannot be previewed. |
| C29 | Government agencies, cities, municipalities and other public sector organizations do not develop their own (open data) activities, and the publication of data must be prescribed by law, and even this often bypasses some, prolongs the waiting times, and often they must be ordered by a court to publish these data. Basically, it is about political culture and what people expect from the public sector. |
| C30 | Government agencies, cities, municipalities and other public sector organizations do not develop their own (open data) activities, and the publication of data must be prescribed by law. |
| C31 | Most of the datasets cannot be visualized, i.e. the visualization feature is not provided or does not work properly for all datasets. |
| D01 | Data governance (governing with data) is preferred as a foundation that enables effective open data arrangements. |
| D02 | The national strategy/policy outline measures to support the reuse of open data by the private sector. |
| D03 | Economic aspects and the overall impact of open data on society and the economy are measured and reported to the audience. |
| D04 | The national strategy/policy outline measures to support the reuse of open data by the public sector. |
| D05 | The open data available on the national portal is accompanied by licensing information. |
| D06 | The portal is reviewed and improved regularly. |
| D07 | The portal's source code as well as relevant documentation and artifacts made available to the public (e.g., on platforms such as GitHub or GitLab). |
| D08 | The users' feedback is considered in the review process of the portal when setting up the agenda for its improvement. |
| D09 | There are regular (at least once a year) business-organized events to improve open data literacy, such as hackathons, workshops, courses, summer/winter schools, user meet-ups. |
| D10 | There are regular (at least once a year) enthusiast-organized (citizens, academia) events to improve open data literacy, such as hackathons, workshops, courses, summer/winter schools, user meet-ups. |
| D11 | There are regular (at least once a year) government-organized events to improve open data literacy, such as hackathons, workshops, courses, summer/winter schools, user meet-ups. |
| D12 | Requests for datasets are processed by representatives of the national open data portal for their compliance with open data principles, evaluate the reliability of their opening and facilitate opening of those meeting these requirements. |
| D13 | Training for administration on opening data to improve the quality and openness of shared data and increase awareness of the benefits of making data available for reuse. |
| D14 | Establish a national AI hub/center to support the public administration in using AI in an ethical, robust, reliable, scalable, and secure manner. |
| D15 | Integration of all types of big data, mostly generated by sensors for publication as open data (in open formats). |
| D16 | Interaction and long-term cooperation with the community and all stakeholders of the ecosystem. |
| D17 | Interoperability and integration are supported to reduce the administrative burden associated with providing services to citizens and business. |
| D18 | Providing guidance and manuals for data providers (ministries, regions, cities, and other stakeholders), i.e., open source software, pattern labs, SEO etc. intended to help other teams create digital products faster (websites and applications) that will be consistent across the public sector agencies and institutions. |
| D19 | Support for allocating and denoting high-value datasets on the OGD portal from technological perspective in accordance with Open Data Directive and requirements ensuring their interoperability. |
| D20 | Support for opening high-value datasets in terms of their determination and preparing for publishing on the OGD portal (in accordance with Open Data Directive and requirements set for their publishing). |
| D21 | Support for the release of valuable datasets (of national value, i.e. country-specific) as open data. |
| D22 | The national portal allows users to request the dataset and track the status of the request in a transparent manner. |
| D23 | The national strategy/policy outline measures to incentivize the publication of and access to citizen-generated data. |
| D24 | The national strategy/policy outline measures to incentivize the publication of and access to geospatial data. |
| D25 | The national strategy/policy outline measures to incentivize the publication of and access to real-time or dynamic data. |